\documentclass[smallextended, pdftex]{svjour3}
\smartqed
\usepackage{graphicx}
\usepackage[utf8]{inputenc}
\usepackage{amsmath,amssymb}
\usepackage{nccmath}
\usepackage{multirow}
\usepackage{here}
\usepackage[absolute]{textpos}
\usepackage{threeparttable}
\usepackage{bm}
\usepackage{lineno,hyperref}
\usepackage{setspace}
\usepackage[square,sort,comma,numbers]{natbib}
\usepackage{subfigure}
\usepackage{comment}
\usepackage{etoolbox}
\usepackage{color}
\newcommand*{\affaddr}[1]{#1}
\newcommand*{\affmark}[1][*]{\textsuperscript{#1}}
\modulolinenumbers[5]

\begin{document}

\title{Detectability of southern gamma-ray sources beyond 100 TeV with ALPAQUITA, the prototype experiment of ALPACA}
\titlerunning{A simulation study on the ALPAQUITA experiment}

\author{S. Kato\affmark[1] \and
C. A. H. Condori\affmark[2] \and
E. de la Fuente\affmark[1, 3, 4, 5] \and
A. Gomi\affmark[6] \and
K. Hibino\affmark[7] \and
N. Hotta\affmark[8] \and
I. Toledano-Juarez\affmark[5, 9] \and
Y. Katayose\affmark[10] \and
C. Kato\affmark[11] \and
K. Kawata\affmark[1] \and
W. Kihara\affmark[11] \and
Y. Ko\affmark[11] \and
T. Koi\affmark[12] \and
H. Kojima\affmark[13] \and
D. Kurashige\affmark[6] \and
J. Lozoya\affmark[14] \and
F. Orozco-Luna\affmark[4] \and
R. Mayta\affmark[15, 16] \and
P. Miranda\affmark[2] \and
K. Munakata\affmark[11] \and
H. Nakada\affmark[6] \and
Y. Nakamura\affmark[1, 17] \and
Y. Nakazawa\affmark[18] \and
C. Nina\affmark[2] \and
M. Nishizawa\affmark[19] \and
S. Ogio\affmark[15, 16] \and
M. Ohnishi\affmark[1] \and
T. Ohura\affmark[6] \and
S. Okukawa\affmark[6] \and
A. Oshima\affmark[12] \and
M. Raljevich\affmark[2] \and
H. Rivera\affmark[2] \and
T. Saito\affmark[20] \and
T. Sako\affmark[1] \and
T. K. Sako\affmark[1] \and
S. Shibata\affmark[12] \and
A. Shiomi\affmark[18] \and
M. Subieta\affmark[2] \and
N. Tajima\affmark[21] \and
W. Takano\affmark[7] \and
M. Takita\affmark[1] \and
Y. Tameda\affmark[22] \and
K. Tanaka\affmark[23] \and
R. Ticona\affmark[2] \and
H. Torres\affmark[24] \and
H. Tsuchiya\affmark[25] \and
Y. Tsunesada\affmark[15, 16] \and
S. Udo\affmark[7] \and
K. Yamazaki\affmark[12] \and
Y. Yokoe\affmark[1]
}

\authorrunning{S. Kato \and}

\institute{
  S. Kato \at \email{katosei@icrr.u-tokyo.ac.jp}\\
  \affaddr{\affmark[1] Institute for Cosmic Ray Research, Kashiwa, 277-8582, Japan}\\
  \affaddr{\affmark[2] Instituto de Investigaciones F\'{i}sicas, Universidad Mayor de San Andres, La Paz 8635, Bolivia}\\
  \affaddr{\affmark[3] Departamento de Fisica, CUCEI, Universidad de Guadalajara, Blvd. Marcelino García Barragán \#1421, esq Calzada Olímpica, C.P. 44430, Guadalajara, Jalisco, Mexico}\\
  \affaddr{\affmark[4] Doctorado en Tecnologías de la Información, CUCEA, Universidad de Guadalajara, Periférico Norte 799, Núcleo Universitario los Belenes, 45100, Zapopan, Jalisco, México}\\
  \affaddr{\affmark[5] Maestría en Ciencias de Datos, CUCEA, Universidad de Guadalajara, Periférico Norte 799 Núcleo Universitario, Los Belenes, 45100 Zapopan, Jalisco, México}\\
  \affaddr{\affmark[6] Graduate School of Engineering Science, Yokohama National University, Yokohama 240-8501, Japan}\\
  \affaddr{\affmark[7] Faculty of Engineering, Kanagawa University, Yokohama 221-8686, Japan}\\
  \affaddr{\affmark[8] Faculty of Education, Utsunomiya University, Utsunomiya 321-8505, Japan}\\
  \affaddr{\affmark[9] Doctorado en Ciencias Fisicas, CUCEI, Universidad de Guadalajara, Blvd. Marcelino García Barragán \#1421, esq Calzada Olímpica, C.P. 44430, Guadalajara, Jalisco, Mexico}\\
  \affaddr{\affmark[10] Faculty of Engineering, Yokohama National University, Yokohama 240-8501, Japan}\\
  \affaddr{\affmark[11] Department of Physics, Shinshu University, Matsumoto 390-8621, Japan}\\
  \affaddr{\affmark[12] College of Engineering, Chubu University, Kasugai, Aichi 487-8501, Japan}\\
  \affaddr{\affmark[13] Chubu Innovative Astronomical Observatory, Chubu University, Kasugai, Aichi 487-8501, Japan}\\
  \affaddr{\affmark[14] Departamento de Ciencias de la Informacion y Desarrollo Tecnologico, Cutonala, Universidad de Guadalajara, Av. 555 Ejido San José Tateposco, Nuevo Perif. Ote., 45425, Tonalá, Jalisco, Mexico}\\
  \affaddr{\affmark[15] Graduate School of Science, Osaka City University, Osaka, Osaka 558-8585, Japan}\\
  \affaddr{\affmark[16] Nambu Yoichiro Institute of Theoretical and Experimental Physics, Osaka City University, Osaka, Osaka 558-8585, Japan}\\
  \affaddr{\affmark[17] Institute of High Energy Physics, Chinese Academy of Science, Shijingsham, Beijing 100049, China}\\
  \affaddr{\affmark[18] College of Industrial Technology, Nihon University, Narashino, 275-8576, Japan}\\
  \affaddr{\affmark[19] National Institute of Informatics, Tokyo 101-8430, Japan}\\
  \affaddr{\affmark[20] Tokyo Metropolitan College of Industrial Technology, Tokyo 116-8523, Japan}\\
  \affaddr{\affmark[21] Institute of Physical and Chemical Research, Wako, 351-0198, Japan}\\
  \affaddr{\affmark[22] Faculty of Engineering, Osaka Electro-Communication University, Neyagawa, Osaka 572-8530, Japan}\\
  \affaddr{\affmark[23] Graduate School of Information Sciences, Hroshima City University, Hiroshima, 731-3194, Japan}\\
  \affaddr{\affmark[24] Coordinación General de Servicios Administrativos e Infraestructura Tecnológica (CGSAIT), Universidad de Guadalajara, Av. Juárez 976, 44100, Guadalajara, Jalisco, México}\\
  \affaddr{\affmark[25] Japan Atomic Energy Agency, Tokai-mura, 319-1195, Japan}\\
}
%\institute{
%  S. Kato \at
%  \email{katosei@icrr.u-tokyo.ac.jp}
%  \and
%}
%\institute{F. Author \at
%              first address \\
%              Tel.: +123-45-678910\\
%              Fax: +123-45-678910\\
%              \email{fauthor@example.com}           %  \\

\date{Received: date / Accepted: date}

\maketitle

\begin{abstract}
  Andes Large-area PArticle detector for Cosmic-ray physics and Astronomy (ALPACA) is an international experiment that applies southern very-high-energy (VHE) gamma-ray astronomy to determine the origin of cosmic rays around the knee energy region ($10^{15}\, {\rm eV} - 10^{16}\, {\rm eV}$). The experiment consists of an air shower (AS) array with a surface of $83,000\, {\rm m}^2$ and an underground water Cherenkov muon detector (MD) array covering $5,400\,{\rm m}^2$. The experimental site is at the Mt.~Chacaltaya plateau in La Paz, Bolivia, with an altitude of $4,740\, {\rm m}$ corresponding to $572\, {\rm g}/{\rm cm}^2$ atmospheric thickness. As the prototype experiment of ALPACA, the ALPAQUITA experiment aims to begin data acquisition in late 2021. The ALPAQUITA array consists of a smaller AS array ($18,450\, {\rm m}^2$) and underground MD ($900\, {\rm m}^2$), which are now under construction. ALPAQUITA's sensitivity to gamma-ray sources is evaluated with Monte Carlo simulations. The simulation finds that five gamma-ray sources observed by H.E.S.S. and HAWC experiments will be detected by ALPAQUITA beyond $10\, {\rm TeV}$ and one out of these five $-$ HESS J1702-420A $-$ above $\simeq 300\, {\rm TeV}$ in one calendar year observation. The latter finding means that scientific discussions can be made on the emission mechanism of gamma rays beyond $100\, {\rm TeV}$ from southern sources on the basis of the observational results of this prototype experiment.
\end{abstract}

\keywords{VHE $\gamma$-ray astronomy \and Monte Carlo simulation \and Detector prototype \and Scientific verification \and TeV cosmic rays}

\section{Introduction}\label{intro}
Cosmic rays (CRs) have been studied since their discovery by Victor Hess in 1912 \cite{Victor_Hess}. The CR energy spectrum is approximately expressed as a power-law spectrum from $10^9\, {\rm eV}$ to $10^{20}\, {\rm eV}$, but it has several bents such as the {\it knee} ($\sim 4\times 10^{15}\, {\rm eV}$) and {\it ankle} ($\sim 10^{18.5}\, {\rm eV}$) \cite{Kristiansen, Bird}. According to theoretical consensus, cosmic rays are of Galactic origin below the knee region. Supernova remnants (SNRs) are the most promising candidates for accelerators because of their shock acceleration \cite{SNR_Bell, CR_SN, Fermi, SNR_Bykov}. Some theories take into account the nonlinear effects of magnetic-field amplification at a shock front to reproduce efficient CR acceleration (for example, see \cite{NLDSA1, NLDSA2}).

Since CRs are deflected by the Galactic magnetic field and lose their directional information, the observation of CRs themselves does not allow us to locate their acceleration sites. One of the effective ways to locate the acceleration sites is to observe gamma rays from the decay of neutral pions produced in the hadronic interaction between CRs and the nearby interstellar medium \cite{Pion_decay_gamma_GeV, Pion_decay_gamma_VHE}. This is because gamma rays are unaffected by the Galactic magnetic field and propagate directly from an acceleration site. Fermi-LAT observed gamma rays from SNR IC443 and reported clear evidence that CRs are accelerated up to ${200\, {\rm GeV}}$ \cite{Fermi_PionDecay}. However, the existence of {\it PeVatrons} that accelerate CRs up to the Galaxy's knee energy region remains to be experimentally verified. Since the neutral-pion-decay gamma rays carry approximately $10\%$ energy from the parent CRs \cite{Kelner}, gamma-ray observation beyond $100\ {\rm TeV}$ is crucial to locate a PeVatron.

Starting with the Crab Nebula, some dozen northern sources have been detected beyond $100\, {\rm TeV}$ by Tibet AS$\gamma$, HAWC and LHAASO \cite{tibet_100TeVCrab, HAWC_100TeVCrab, LHAASO_100TeVCrab, HAWC_56TeV, LHAASO_100TeV}, and some of the sources such as G106.3+2.7 and the Cygnus OB1 and OB2 associations are found to be promising candidates for PeVatron \cite{HAWC_G106, Tibet_G106, HAWC_Cygnus, Tibet_Cygnus}. Tibet AS$\gamma$ has also found that PeVatrons currently exist or at least existed in the Galaxy by observing Galactic diffuse gamma rays up to $1\, {\rm PeV}$ \cite{Tibet_diffuse}. However, current experiments sensitive to gamma rays beyond $100\, {\rm TeV}$ are located only in the northern hemisphere and do not have access to the southern sky, where about 100 VHE gamma-ray sources, including the Galactic Center, have been detected \cite{HESS_PeVatron, HESS_GC2, HGPS}. In other words, gamma-ray observation beyond $100\, {\rm TeV}$ in the southern sky remains lacking, in spite of its essential importance in terms of the identification of PeVatrons and the distribution of cosmic rays around the knee energy region \cite{Lipari_diffuse}. Under these circumstances, it is crucial to design an experiment sensitive to gamma rays beyond $100\, {\rm TeV}$ in the southern sky.

This research proceeds with a new air shower array experiment in the southern hemisphere, the ALPACA experiment. ALPACA aims to determine the origin of CRs around the knee energy region through 100 TeV gamma-ray observation. ALPAQUITA is designed as the prototype experiment of ALPACA. It is fruitful to evaluate the possibility of $100\, {\rm TeV}$ gamma-ray detection from celestial sources with ALPAQUITA. This will help us to explore southern gamma-ray sky beyond $100\, {\rm TeV}$ which current experiments do not have access to.

Using a Monte Carlo simulation, this paper discusses the expected performance of ALPAQUITA as a gamma-ray observatory, including its sensitivity to gamma-ray point sources. Section \ref{experiment} introduces the ALPACA and ALPAQUITA experiments. The simulation configuration is presented in Section \ref{MCsimulation}. Sections \ref{MCanalysis} and \ref{MDanalysis} describe the event reconstruction methods and the resultant performance, respectively. Section \ref{Conclusion} presents conclusions and directions for future research.

%\paragraph{Installation} If the document class \emph{elsarticle} is not available on your computer, you can download and install the system package \emph{texlive-publishers} (Linux) or install the \LaTeX\ package \emph{elsarticle} using the package manager of your \TeX\ installation, which is typically \TeX\ Live or Mik\TeX.

%\paragraph{Usage} Once the package is properly installed, you can use the document class \emph{elsarticle} to create a manuscript. Please make sure that your manuscript follows the guidelines in the Guide for Authors of the relevant journal. It is not necessary to typeset your manuscript in exactly the same way as an article, unless you are submitting to a camera-ready copy (CRC) journal.

%\paragraph{Functionality} The Elsevier article class is based on the standard article class and supports almost all of the functionality of that class. In addition, it features commands and options to format the
%\begin{itemize}
%\item document style
%\item baselineskip
%\item front matter
%\item keywords and MSC codes
%\item theorems, definitions and proofs
%\item lables of enumerations
%\item citation style and labeling.
%\end{itemize}

\section{The ALPACA and ALPAQUITA experiments}\label{experiment}
\subsection{The ALPACA experiment}\label{ALPACA}
ALPACA is an international experiment among Bolivia, Mexico, and Japan. The experimental site is located at Chacaltaya Plateau ($4,740\, {\rm m}$ a.s.l., $16^{\circ}\, 23^{'}\,{\rm S}, $\\ $68^{\circ}\, 08^{'}\, {\rm W}$), halfway up Mt.~Chacaltaya, Bolivia. The altitude corresponds to $572\, {\rm g}/{\rm cm}^2$ atmospheric thickness, near which air showers generated by $100\, {\rm TeV}$ gamma rays reach their maximum development.

ALPACA consists of a surface air shower (AS) array and an underground water Cherenkov muon detector (MD) array. The total areas of the AS and MD arrays are $83,000\, {\rm m}^2$ and $5,400 \, {\rm m}^2$, respectively. ALPACA will cover various topics in cosmic-ray physics, including the origin and chemical composition of CRs around the knee energy region, the anisotropy of CRs in the TeV-PeV range, and the modeling of the interplanetary magnetic field using the Sun shadow in cosmic rays \cite{Tibet_Composition, Tibet_Aniso, Tibet_SMF, Tibet_IMF}.

\subsection{The ALPAQUITA experiment}\label{ALPAQUITA}
\begin{figure}[h]
  \begin{center}
    \includegraphics[scale=0.2]{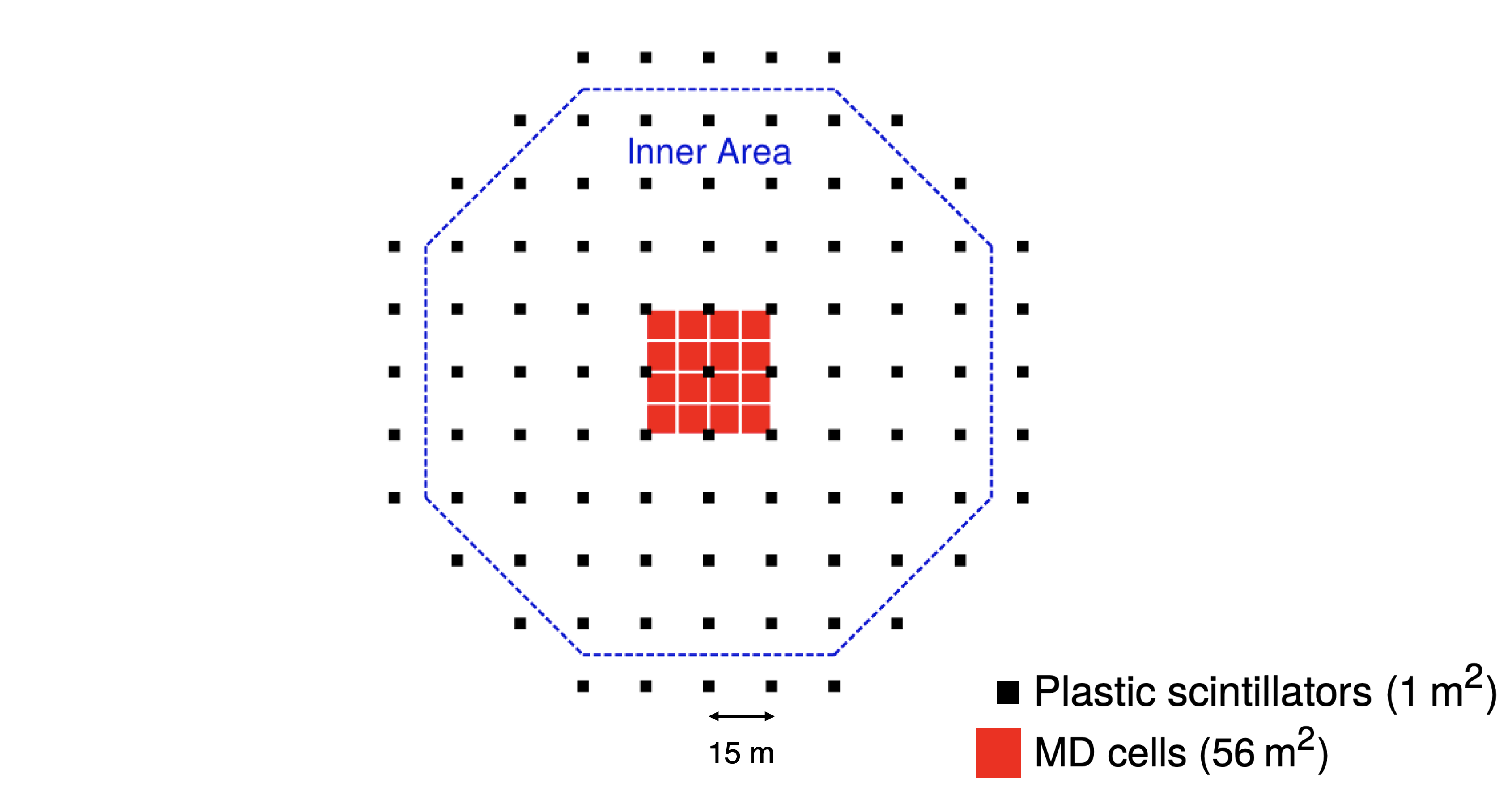}
  \end{center}
  \caption{Schematic view of ALPAQUITA. Black squares indicate the plastic scintillation detectors, each of which has an area of $1\, {\rm m}^2$. Red squares indicate the muon detector (MD) cells, each of which has an area of $56\, {\rm m}^2$. Regarding the inner area shown by dashed blue lines, see Section \ref{Ana_conditions}
  }
  \label{ALPAQUITA_schematic}
\end{figure}

ALPAQUITA is the prototype experiment of ALPACA, and its array is now under construction. The experiment aims to start data acquisition in late 2021 and, after short observation, expand the array to the ALPACA array. Figure \ref{ALPAQUITA_schematic} shows the schematic view of ALPAQUITA. The AS array comprises 97 plastic scintillation detectors, each of which has an area of $1\, {\rm m}^2$ and is located at intervals of $15\, {\rm m}$ in a grid pattern. The total area is $18,450\, {\rm m}^2$, a quarter the size of the ALPACA AS array. The AS array is used to trigger shower events and reconstruct the energy and incoming direction of a primary particle, as presented in Section \ref{reconstruction}.

\begin{figure}[h]
  \begin{center}
    \includegraphics[scale=0.2]{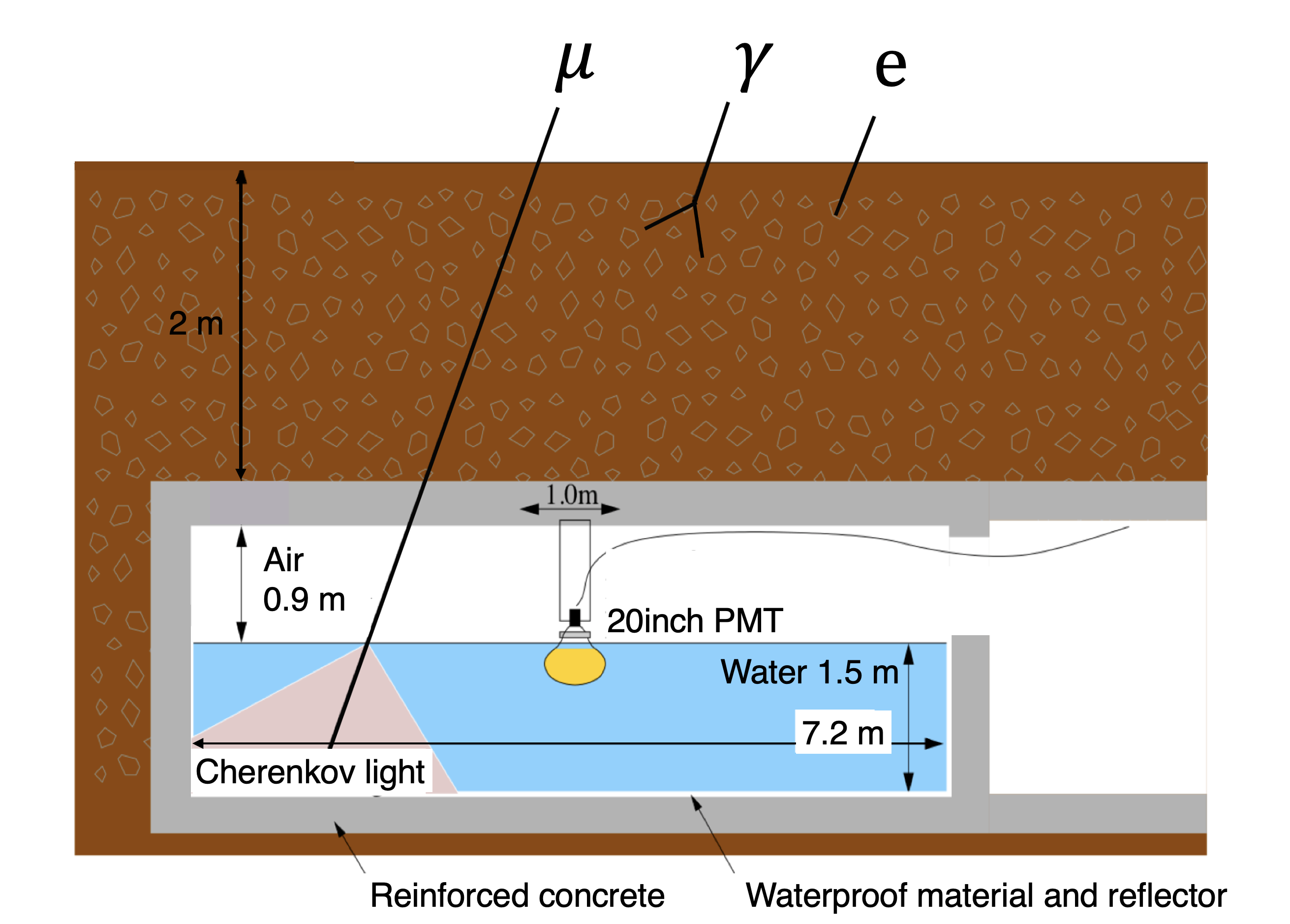}
  \end{center}
  \caption{Schematic view of the MD cell}
  \label{MDcell_schematic}
\end{figure}

The ALPAQUITA's MD consists of 16 smaller cells and has a total area of $900\, {\rm m}^{2}$ (see Figure \ref{ALPAQUITA_schematic}). Figure \ref{MDcell_schematic} shows the schematic view of the cells. The total thickness of the soil above MD and the concrete ceiling corresponds to $17$ radiation lengths. Therefore, most of the electromagnetic components in air showers are absorbed in the soil and concrete layers, while only muons with energies $\gtrsim 1.2\, {\rm GeV}$ can reach MD and emit Cherenkov light in the water layer. The Cherenkov light is collected with a photomultiplier tube (PMT) installed downward at the ceiling. Since CR-induced air showers are muon-rich while gamma-ray-induced are muon-poor, MD can efficiently discriminate primary gamma rays from CRs.

\section{Monte Carlo simulation}\label{MCsimulation}
%In this section, we introduce the simulation configuration employed to evaluate the ALPAQUITA performance. We have two parts to be simulated; an air shower simulation and a detector-response simulation. We describe these two parts in order.

\subsection{Air shower simulation}\label{CORSIKA}
Primary gamma-ray and CR events are generated and air shower development is simulated with CORSIKA7.6400 \cite{corsika}. Primary gamma rays are generated following a simple power-law spectrum with an index of 2 within the energy range of $300\, {\rm GeV}\le E \le 10\, {\rm PeV}$ from a hypothetical point source along the path in the sky of RX J1713.7-3946, a bright gamma-ray source in the southern sky \cite{Cangaroo, FermiLAT_RXJ1713, RXJ1713}. Assuming this path, the minimum zenith angle is $23.4^{\circ}$ at the ALPAQUITA site. In this simulation, the injection's zenith angle is limited within $60^{\circ}$, beyond which the detection efficiency of 100 TeV gamma rays becomes very low. Along the aforementioned path in the sky, $3.7 \times 10^{7}$ gamma-ray events are generated and injected into the atmosphere where the air shower development is simulated. As a simulation area, a circular region with a $300\, {\rm m}$ radius from the center of the ALPAQUITA AS array is assumed, and shower cores are randomly distributed within the area. The generated gamma-ray events are weighted to modify the index and normalization of the spectrum depending on the analysis.

Primary CR events are also generated along the RX J1713.7-3946 path in the sky and injected into the atmosphere under the same conditions as gamma-ray events. The modification of the number of CR events considering the isotropic characteristics is made in Monte Carlo data analysis procedures (see Section \ref{MD_ana}). For the chemical composition and energy spectrum, the model spectrum proposed by M. Shibata et al. (2010) \cite{MixPrimary} is adopted. FLUKA \cite{FLUKA} and EPOS LHC \cite{EPOS-LHC} are employed as the low- and high-energy hadronic interaction models, respectively. $8.5 \times 10^{8}$ events are generated in the energy range of $300\, {\rm GeV} \le E \le 10\, {\rm PeV}$ and additional $7.7 \times 10^{7}$ events in $10\, {\rm TeV} \le E \le 10\, {\rm PeV}$ to increase statistics in the high-energy range. They correspond to $\simeq 0.4$ and $\simeq 5$ years of statistics.

\subsection{Detector simulation}\label{G4}
Based on GEANT4.v10.04.p02 \cite{Geant4}, detector responses of the AS array and MD to the shower events are simulated. Simulation settings for the AS array and MD are separately described in this section, and parameters used in the detector simulations are summarized in Table \ref{Scinti_table}.

\begin{figure}[h]
  \begin{center}
    \includegraphics[scale=0.25]{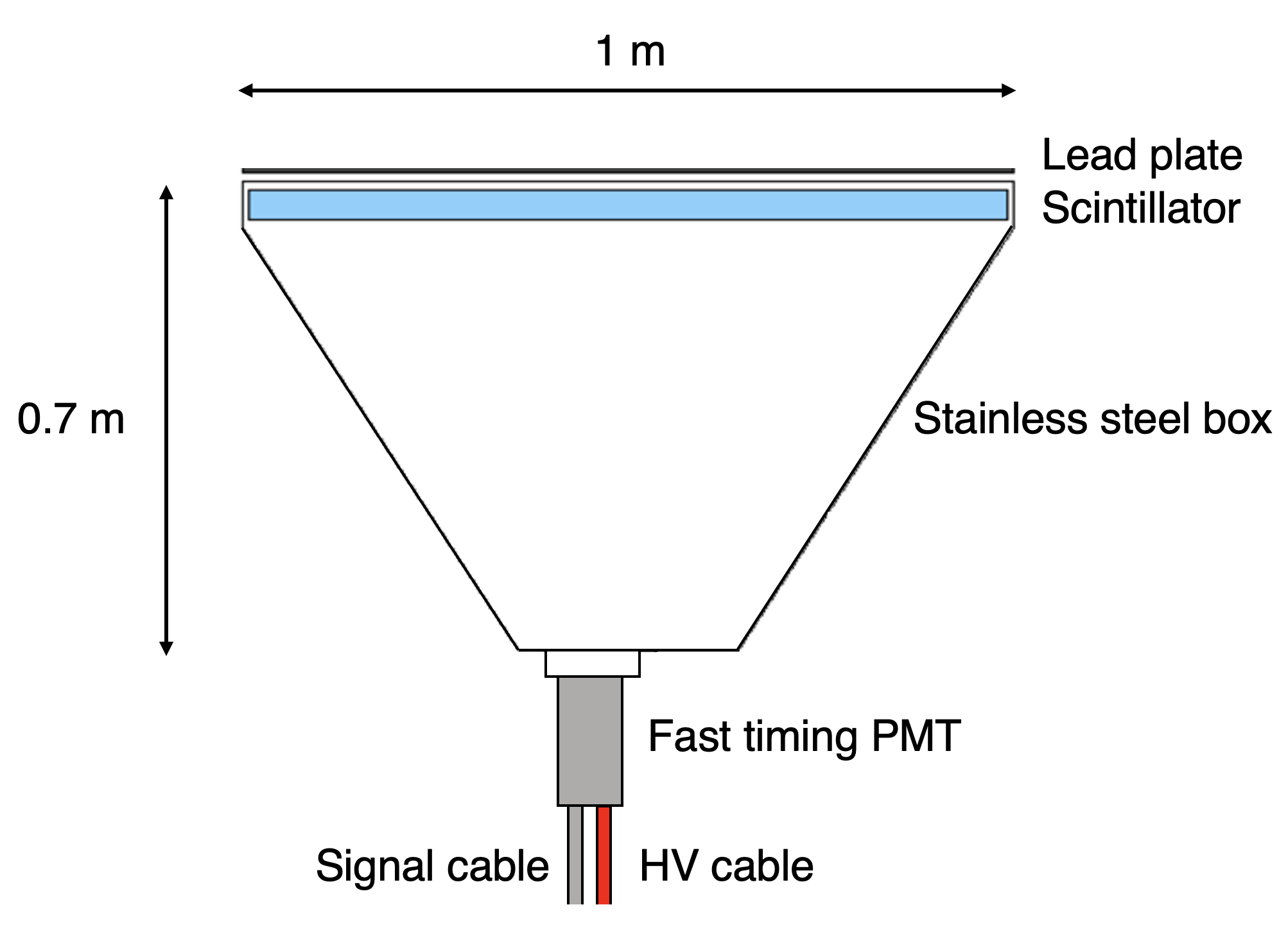}
  \end{center}
  \caption{Schematic view of the plastic scintillation detector. For a detailed description, see Section \ref{G4}}
  \label{Scintillator}
\end{figure}

\paragraph{AS-array configuration} Figure \ref{ALPAQUITA_schematic} presents the configuration of the ALPAQUITA AS array. A total of 97 plastic scintillation detectors are located with $15\, {\rm m}$ spacing and cover a total area of $18,450\, {\rm m}^2$. Figure \ref{Scintillator} shows the design of each scintillation detector. A plastic scintillator of $1\, {\rm m}^2$ area and $5\, {\rm cm}$ thickness is installed on the top of a steel box with a lead plate of $5\, {\rm mm}$ thickness on it. This plate is used to induce $e^{\pm}$ pair productions of secondary gamma rays in an air shower and double energy deposit in the plastic scintillator. A fast timing PMT is installed at the bottom of the steel box and collects the scintillation light diffusely reflected inside the box. The energy deposit of shower particles in the scintillation detectors is calculated and converted into the number of particles where the single-particle peak is defined as $9.4\, {\rm MeV}$. Triggers are issued when any four scintillation detectors detect more than $0.5$ particles within $600\, {\rm ns}$ \cite{TibetMD}.

\begin{figure}[h]
  \begin{center}
    \includegraphics[scale=0.4]{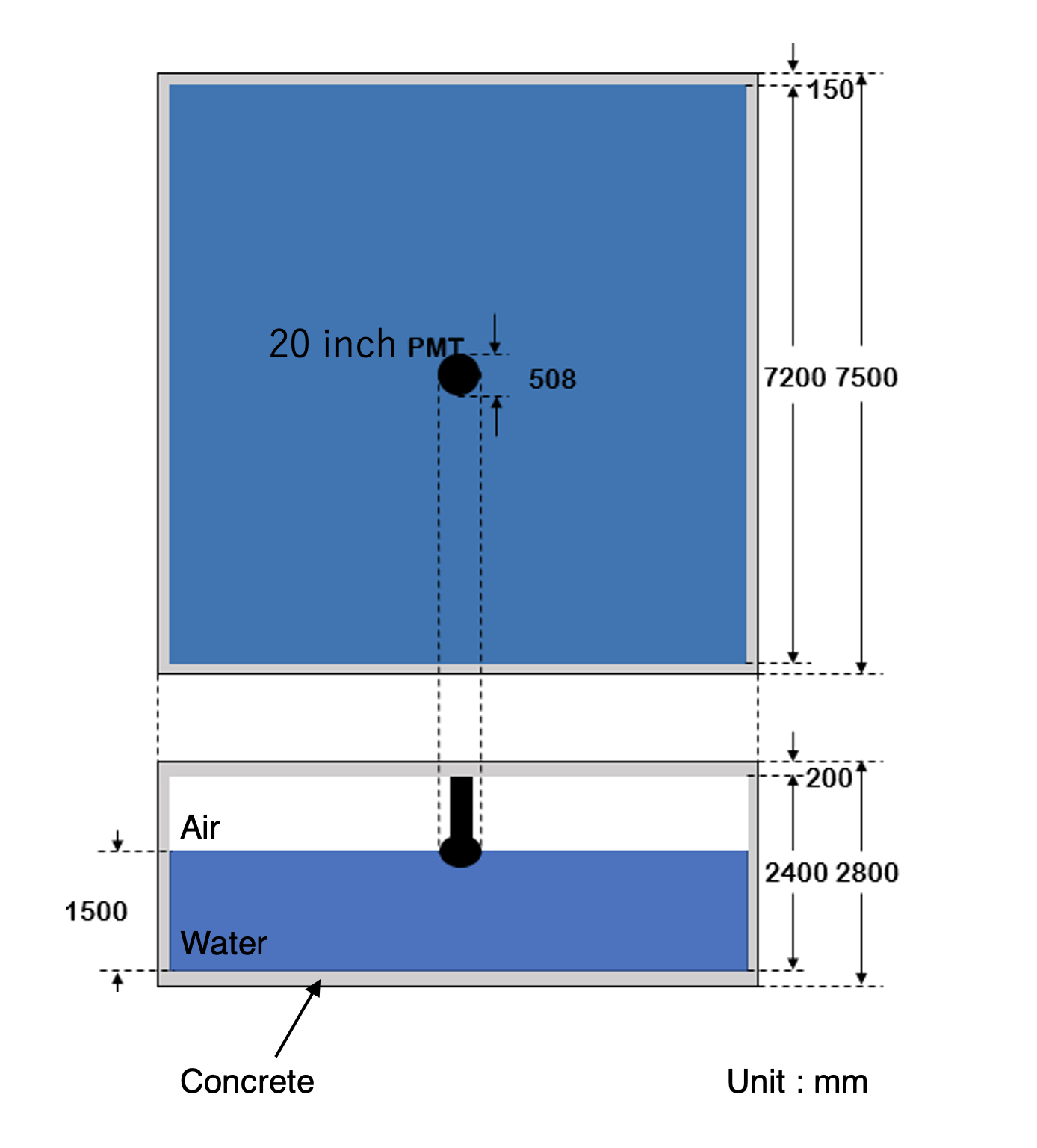}
  \end{center}
  \caption{Design of the MD cell. For a detailed description, see Section \ref{G4}}
  \label{MD_design}
\end{figure}

\paragraph{MD configuration} MD has a total area of $900\, {\rm m}^2$ and is located beneath a soil overburden of $2\, {\rm m}$ at the AS array center. The detector consists of 16 cell units, each of which is designed as shown in Figure \ref{MD_design}. The cells are made of reinforced concrete and contain a water layer of $1.5\, {\rm m}$ thickness with an air layer of $0.9\, {\rm m}$ thickness. A 20-inch PMT is suspended downward at the center of the ceiling, and its photo-sensitive area is located below the water surface. For events that trigger the AS array, a trigger gate is opened for MD and the PMT signals are calculated in the unit of the number of photoelectrons by simulating Cherenkov light emission and the paths of the light in the water layer for the particles that reach MD. The Cherenkov light is diffusely reflected from the cell wall in the water layer with a reflectivity of $80\%$. The quantum efficiency of the photocathode is assumed to be the same as that adopted in Tibet AS$\gamma$ \cite{TibetMD}. The first-dynode collection efficiency and the geomagnetic effect on the number of collected photoelectrons are also considered.

There may be accidental muons that are irrelevant to triggered shower events and distributed uniformly over the detector. The number of accidental muons which contaminate the MD signals is assumed to follow a Poisson distribution with mean $0.18$ over MD per triggered event. This mean value is based on the results of Tibet AS$\gamma$ \cite{TibetMD}.
\clearpage
%\begin{textblock*}{0.4\linewidth}(40pt, 0pt)
\begin{table}[H]
%  \begin{center}
  \hspace{-1.5cm}
    \begin{threeparttable}
    \caption{Parameter settings of the plastic scintillation detectors and the MD cells.}
    \begin{tabular}{cccc} \hline \hline
      Part & Material\tnote{1} & Density\tnote{2} & Size\tnote{3} \\ \hline
      %        & & (${\rm g}/{\rm cm}^3$) & (length$({\rm cm})\times$width(${\rm cm}$)$\times$ thickness(${\rm cm}$))\\ \hline
      %info @ geant/source/materials/src/G4NistMaterialBuilder.cc
      Atmosphere & ${\rm N}\, (75.5\%)$, ${\rm O}\, (23.2\%)$, ${\rm Ar}\, (1.3\%)$ and ${\rm C}\, (23.2\%)$ & 572.4 & -\\
      Soil & ${\rm SiO_2}\, (70\%)$, ${\rm Al_2O_3}\, (20\%)$, and ${\rm CaO}\, (10\%)$ & 2.1 & -\\
      Plastic scintillator & ${\rm C}_{9}{\rm H}_{10}$ & 1.032 & 100$\times$100$\times$5\\
      Steel box & ${\rm Fe}\, (74\%)$, ${\rm Cr}\, (18\%)$, ${\rm Ni}\, (8\%)$ & 7.820 & 103.3$\times$103.3$\times$0.1\\
      Lead plate & Pb & 11.34 & 100$\times$100$\times$0.5\\
      \multirow{2}{*}{MD cells (Reinforced concrete)} & ${\rm H}\, (1.0\%)$, ${\rm C}\, (0.1\%)$, ${\rm O}\, (52.9\%)$, ${\rm Na}\, (1.6\%)$, ${\rm Mg}\, (0.2\%)$, & \multirow{2}{*}{2.3} & \multirow{2}{*}{See Figure \ref{MD_design}}\\
      & ${\rm Al}\, (3.4\%)$, ${\rm Si}\, (33.7\%)$, ${\rm K}\, (1.3\%)$, ${\rm Ca}\, (4.4\%)$, and ${\rm Fe}\, (1.4\%)$& &\\ \hline
    \end{tabular}
    \label{Scinti_table}
    \begin{tablenotes}
      \begin{spacing}{0.9}
      \item[1] All the percentages indicate mass fraction.
      \item[2] The unit is ${\rm g}/{\rm cm}^3$ except for the atmosphere (${\rm g}/{\rm cm}^2$).
      \item[3] Length (${\rm cm}$) $\, \times$ width (${\rm cm}$) $\, \times$ thickness (${\rm cm}$).
      \end{spacing}
    \end{tablenotes}
  \end{threeparttable}
%  \end{center}
  \end{table}
%\end{textblock*}

\section{Monte Carlo data analysis}\label{MCanalysis}
Using the simulation data, the performance of the ALPAQUITA AS array for gamma rays is evaluated. Performance for gamma-hadron separation is discussed in Section \ref{MDanalysis}.

\subsection{Electronics settings}\label{El_set}
The obtained simulation data are converted into binary format and analyzed in the same way as experimental data. The ADC charge resolution is assumed as $0.076\, {\rm pC/count}$. For the ADC counts of each scintillation detector, random number counts are added, following a Gaussian with mean $0$ and standard deviation $2$ to take into account electric noise. The single-particle peak of each detector is assumed as $17\, {\rm pC}$. The TDC resolution is assumed as $0.5\, {\rm ns/count}$, and electric fluctuation following Gaussian fluctuation with mean $0\, {\rm ns}$ and standard deviation $0.2\, {\rm ns}$ is mixed into the TDC counts of each detector.

\subsection{Reconstruction of primary information}\label{reconstruction}
To estimate the energies of primary particles, the correlation between primary energy and the number of particles detected with the AS array is analyzed. For this purpose, $\Sigma \rho$ is defined as the sum of the density of detected particles  $\rho$  over all the hit detectors except for the detector that records the largest particle density and is used as an energy estimator. The procedure of removing the largest contribution is because such a contribution is largely affected by air shower development fluctuation and worsens the resultant energy resolution. The conversion function from $\Sigma \rho$ to primary energy for selected gamma-ray events is discussed in Section \ref{ALPAQUITA_AS_prop}.

A shower core position $(x_{\rm c}, \, y_{\rm c})$ is assumed by the weighted average of the positions of surface scintillation detectors as
\begin{equation}\label{core_estimation}
  (x_{\rm c}, \, y_{\rm c}) = \bigg(\frac{\sum_i {\rho_i}^2 x_i}{\sum_i {\rho_i}^2}, \, \frac{\sum_i {\rho_i}^2 y_i}{\sum_i {\rho_i}^2} \bigg),
\end{equation}
where $x_i$ and $y_i$ are the coordinates of the $i$-th scintillation detector. The summation runs over all the hit detectors that record more than $0.8$ particles.

\begin{figure}[h]
  \begin{center}
    \includegraphics[scale=0.3]{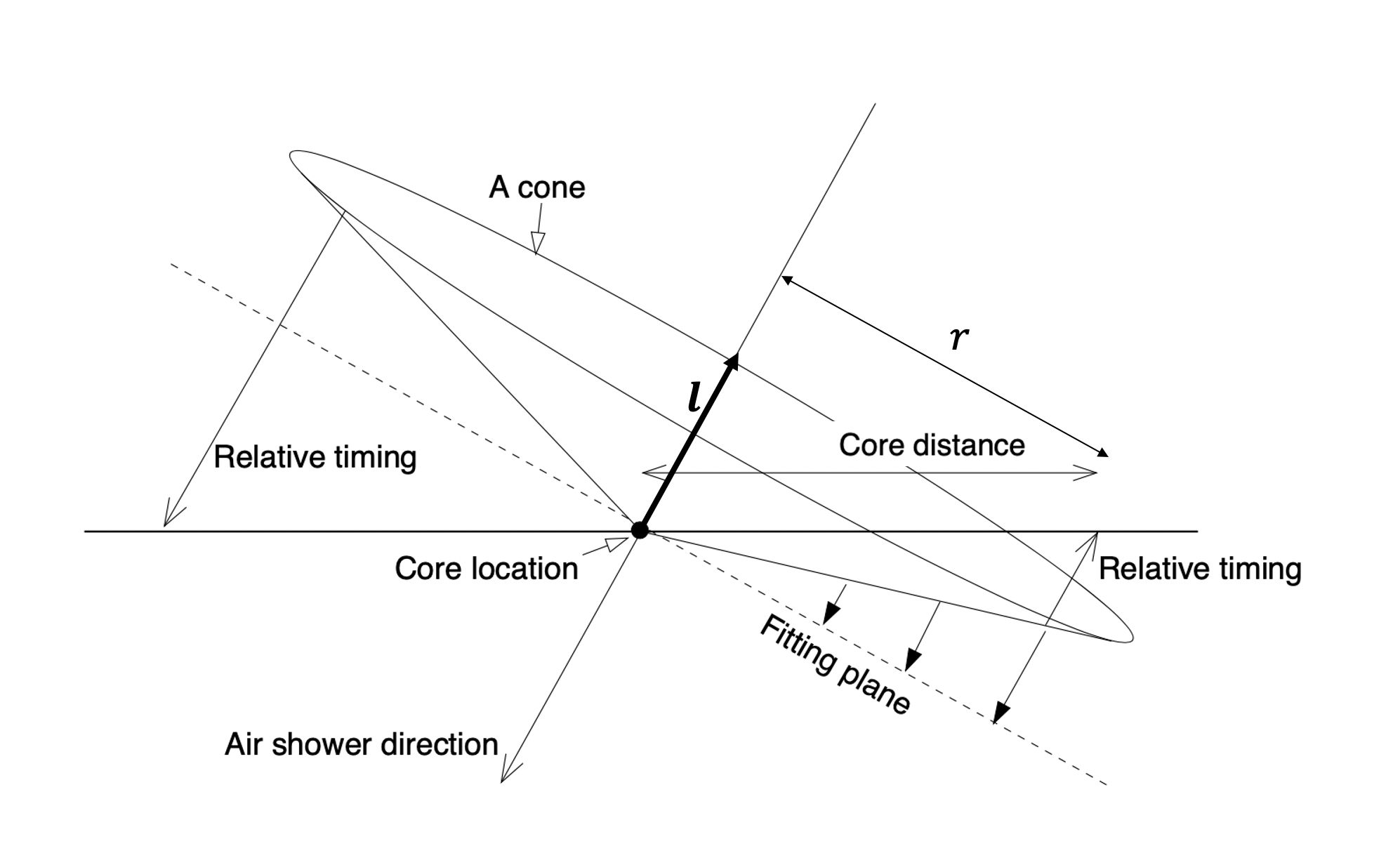}
  \end{center}
  \caption{Principle of the method that determines the incoming direction of a shower event. For a detailed description of the reconstruction method, see Section \ref{reconstruction}}
  \label{Residual}
\end{figure}

For the estimation of arrival direction, the relative timing information of the hit detectors is used. Figure \ref{Residual} shows a schematic view of the estimation procedure. A shower front is assumed in a conical shape having a slope
\begin{equation}\label{cone_fit}
  b \, ({\rm ns}/{\rm m}) = 0.0125 \, {\rm log}_{10} \bigg(\frac{\sum_i \rho_i}{{\rm m}^{-2}}\bigg) + 0.0625, \ \ \ \ 0.075\le b \le 0.12
\end{equation}
where the summation is taken over all the hit detectors, and $\rho_i$ is fixed to $45\, {\rm m}^{-2}$ for $\rho_i \ge 45\, {\rm m}^{-2}$. The relative hit timing of the $i$-th detector $t_i$ is modified as
\begin{equation}\label{modulation}
 t^{'}_{i} = \, t_i - b \, r,  \nonumber
\end{equation}
depending on the distance $r$ between the detector and the shower axis which corresponds to the directional cosine $\bm{l}$ of a shower event. After the modification, $\bm{l}$ is calculated, which minimizes the square of $\chi$, called residual error:
\begin{eqnarray}\label{residual}
  \chi^2\, = \sum_i w_i (\bm{l} \cdot \bm{x}_{i} + c ({t}^{'}_{i} - t_0))^2,\,\,\,\,\,\,
  (w_i = \rho_i/{\Sigma_i \rho_i}, \ \bm{x}_i = (x_i,\, y_i,\, 0))
\end{eqnarray}
where $c$ is the speed of light and the summation is taken over all the hit detectors, and $\rho_i$ is fixed to $30\, {\rm m}^{-2}$ for $\rho_i \ge 30\, {\rm m}^{-2}$. $\chi$ indicates the quality of reconstruction (i.e., if the residual error is large for a recorded event, then the incoming direction of the event is poorly reconstructed.) The above procedures are iterated several times until $\bm{l}$ converges, determining the incoming direction of the primary particle.

\subsection{Analysis conditions}\label{Ana_conditions}
For reconstructed events, several analysis conditions are imposed. For the analysis of the AS array performance, the following three conditions are employed: (1) any four scintillation detectors detect more than 0.8 particles, (2) three out of the four detectors that record the largest particle densities are inside the inner area, which is surrounded by dashed blue lines in Figure \ref{ALPAQUITA_schematic}, and (3) the residual error $\chi$ calculated by Equation \eqref{residual} is smaller than $1.0\, {\rm m}$. 
  
\subsection{Performance of the ALPAQUITA AS array for primary gamma rays}\label{ALPAQUITA_AS_prop}
This section presents the AS array performance for gamma rays: trigger efficiency, energy resolution, the accuracy of shower core position estimation, and angular resolution. Note that the performance is evaluated for gamma rays that follow a simple power-law spectrum with an index of $2.5$ and have a true zenith angle within $40^{\circ}$.

\begin{figure}[h]
  \begin{center}
    \includegraphics[scale=0.25]{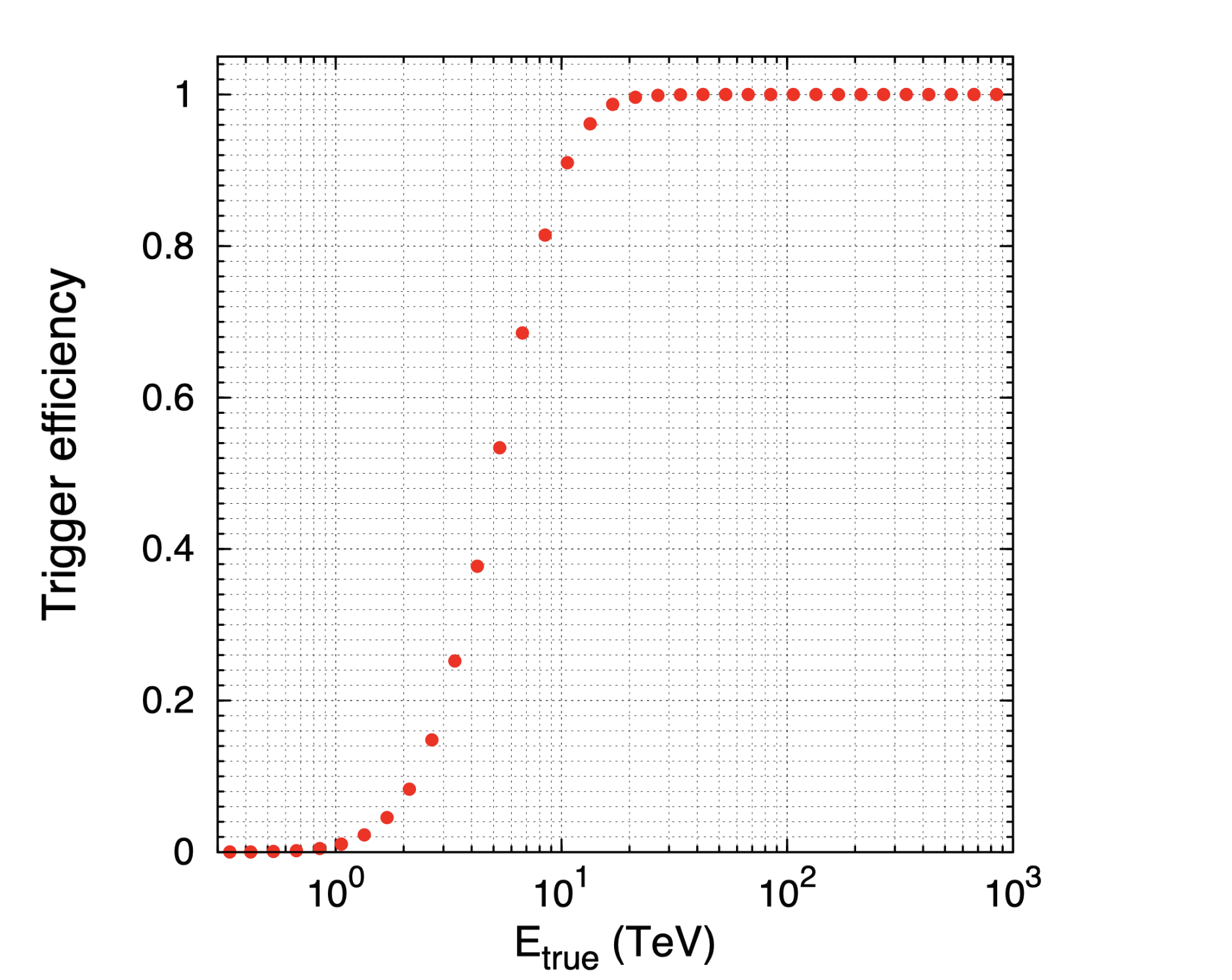}
  \end{center}
  \caption{Estimated trigger efficiency of the ALPAQUITA AS array as a function of the true energies of primary gamma rays}
  \label{Trig_effi}
\end{figure}

Figure \ref{Trig_effi} shows the trigger efficiency of the ALPAQUITA AS array as a function of the true energies of primary gamma rays $E_{\rm true}$. The trigger efficiency is defined for the events that have the true shower core positions inside the ALPAQUITA AS array and pass the trigger condition presented in Section \ref{G4}. The efficiency is estimated to reach 100\% above $20\, {\rm TeV}$.

\begin{figure}[h]
  \begin{center}
    \includegraphics[scale=0.25]{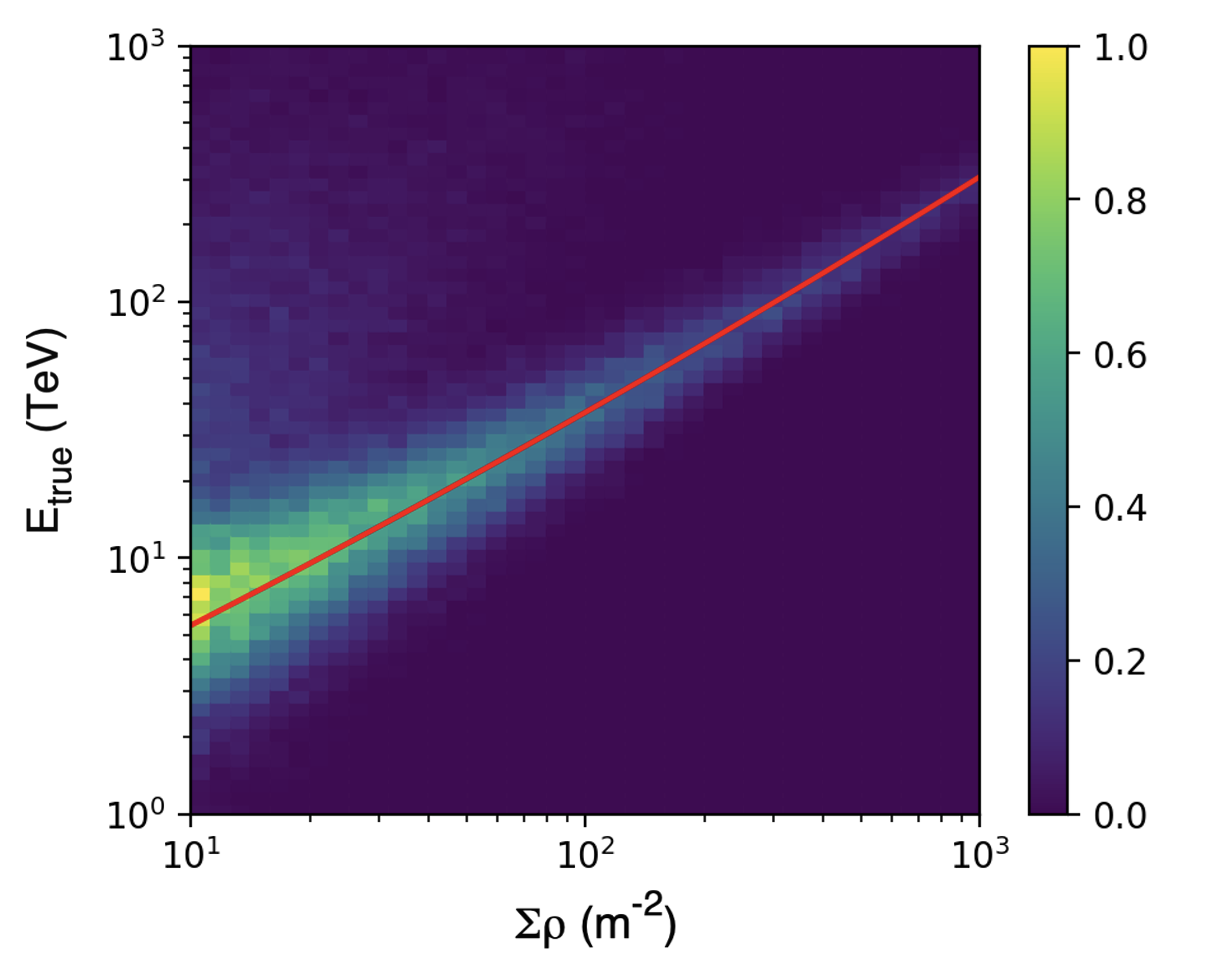}
  \end{center}
  \caption{Relation between $\Sigma \rho$ and $E_{\rm true}$. The red curve shows the conversion function from $\Sigma \rho$ to gamma-ray energy employed in the analysis (see Section \ref{ALPAQUITA_AS_prop})}
  \label{srho_ene_scat}
\end{figure}

\begin{figure}[h]
  \begin{center}
    \includegraphics[scale=0.5]{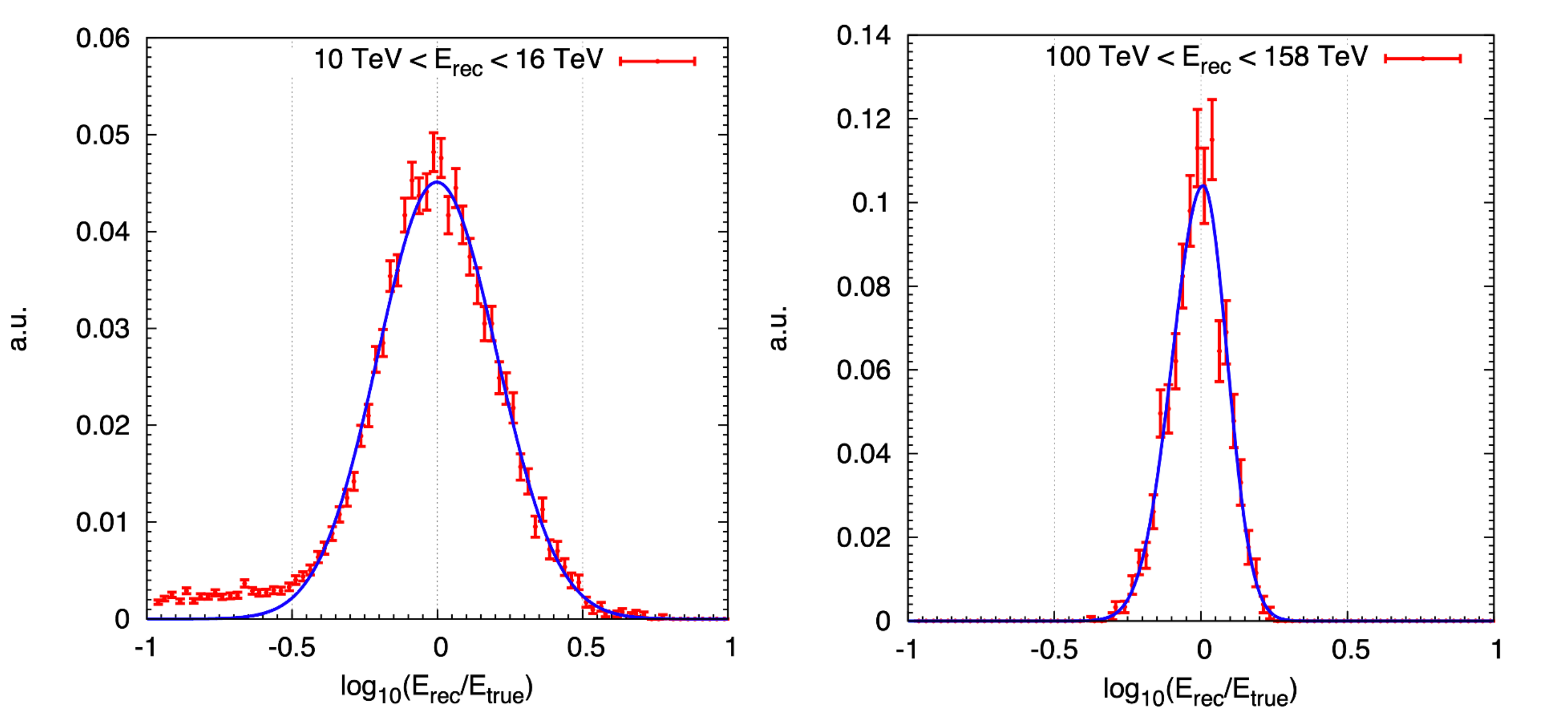}
  \end{center}
  \caption{{\it Left}: Distribution of the logarithm of the ratio of the reconstructed energy to the true one (red) of primay gamma rays in the reconstructed energy ($E_{\rm rec}$) range of $10\, {\rm TeV} < E_{\rm rec} < 16\, {\rm TeV}$. The blue curve shows the asymmetric Gaussian fitted to the distribution. {\it Light}: The same distribution in $100\, {\rm TeV} < E_{\rm rec} < 158\, {\rm TeV}$}
  \label{Rep_ene}
\end{figure}

\begin{figure}[h]
  \begin{center}
    \includegraphics[scale=0.6]{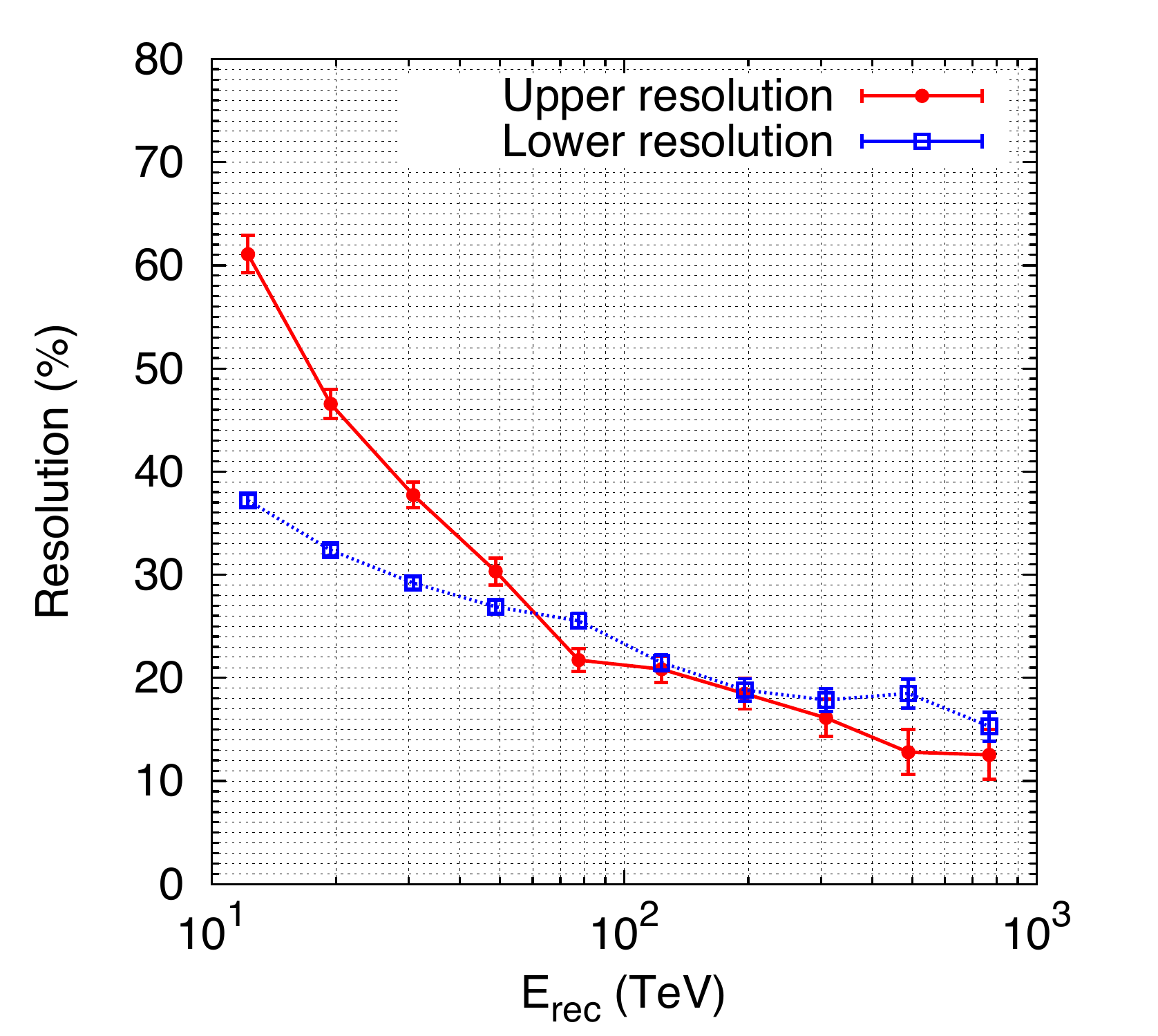}
  \end{center}
  \caption{Upper (red) and lower (blue) energy resolutions of the ALPAQUITA AS array for primary gamma rays as a function of reconstructed energy. For the definition of the upper and lower resolutions, see Section \ref{ALPAQUITA_AS_prop}}
  \label{ene_resolution}
\end{figure}

The relation between the true energies of primary gamma rays and $\Sigma \rho$ is shown in Figure \ref{srho_ene_scat}. The red curve shows the optimum energy conversion formula and is written as
\begin{equation}\label{srho_energy_relation}
  {\rm log}_{10}\, \Big(\frac{E}{\rm TeV}\Big) = 4.4\times 10^{-2}\bigg({\rm log}_{10}\, \Big(\frac{\Sigma \rho}{{\rm m}^{-2}}\Big)\bigg)^{2} + 7.0\times 10^{-1}\bigg({\rm log}_{10}\,\Big(\frac{\Sigma \rho}{{\rm m}^{-2}}\Big)\bigg) - 7.6\times 10^{-3}. \nonumber
\end{equation}
Applying the formula to the analysis, the distribution of the logarithm of the ratio of the reconstructed energy to the true one of primary gamma rays is calculated for each reconstructed energy bin as shown in Figure \ref{Rep_ene}. The distribution is fitted with the asymmetric Gaussian, and the upper (lower) energy resolution is defined as the right (left) deviation of 1 $\sigma$ from the peak of the distribution. Figure \ref{ene_resolution} shows the resultant resolutions, and the upper and lower energy resolutions are empirically formulated as $\simeq 21\, (0.5\, {\rm log}_{10}(E_{\rm rec}/100\, {\rm TeV}) + 1)^{-1.5}\, \%$ and $\simeq 23\, (0.5\, {\rm log}_{10}(E_{\rm rec}/100\, {\rm TeV}) + 1)^{-0.7}\, \%$ in $10\, {\rm TeV} < E_{\rm rec} < 1000\, {\rm TeV}$, respectively.

\begin{figure}[h]
  \begin{center}
    \includegraphics[scale=0.5]{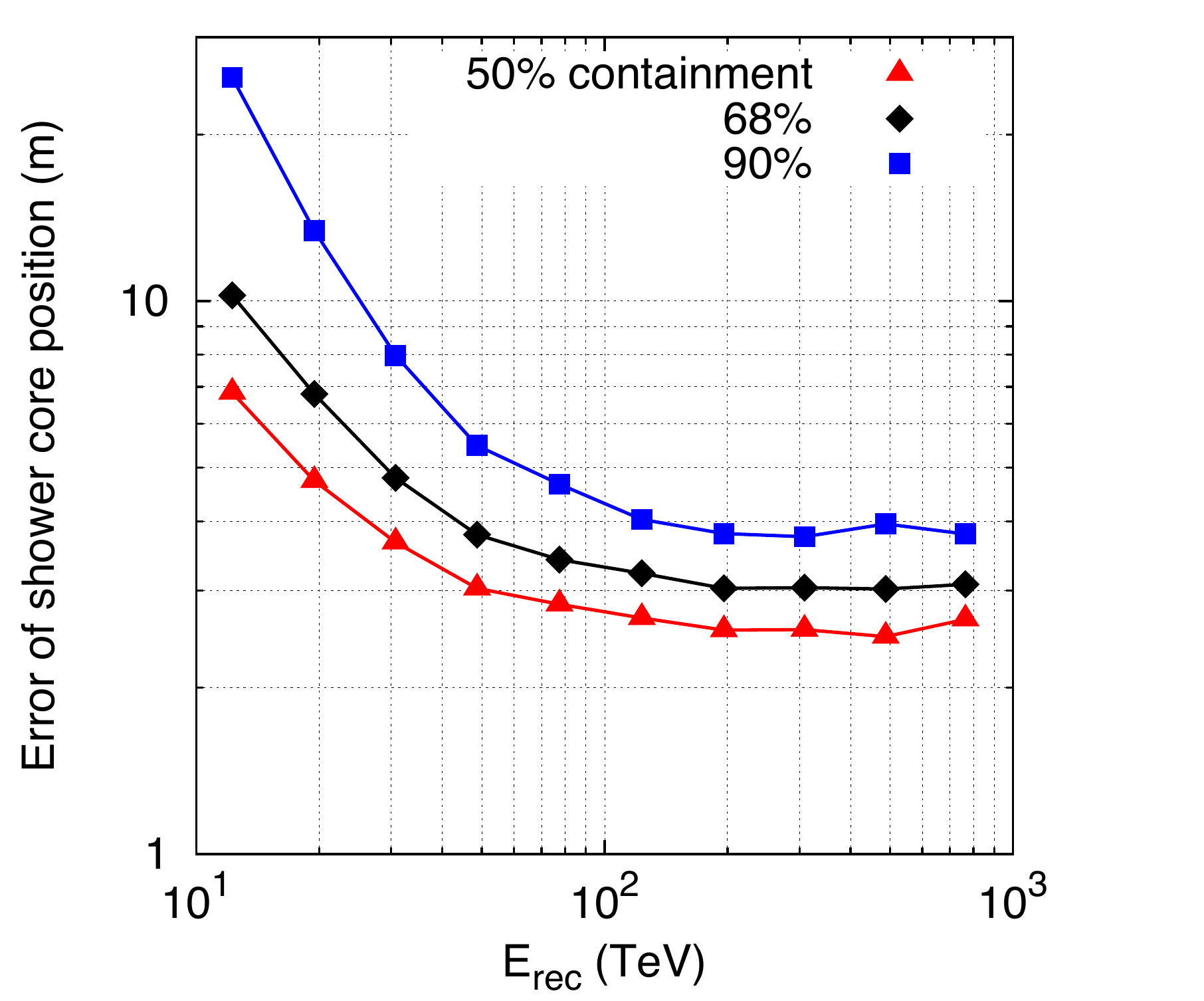}
  \end{center}
  \caption{Shower core position resolution (50\% containment, red triangle) of the ALPAQUITA AS array for gamma-ray induced air shower events as a function of reconstructed energy. Also shown are the angular radii inside which 68\% (black diamond) and 90\% (blue square) of analyzed gamma-ray events are contained}
  \label{Core_res}
\end{figure}

\begin{figure}[h]
  \begin{center}
    \includegraphics[scale=0.5]{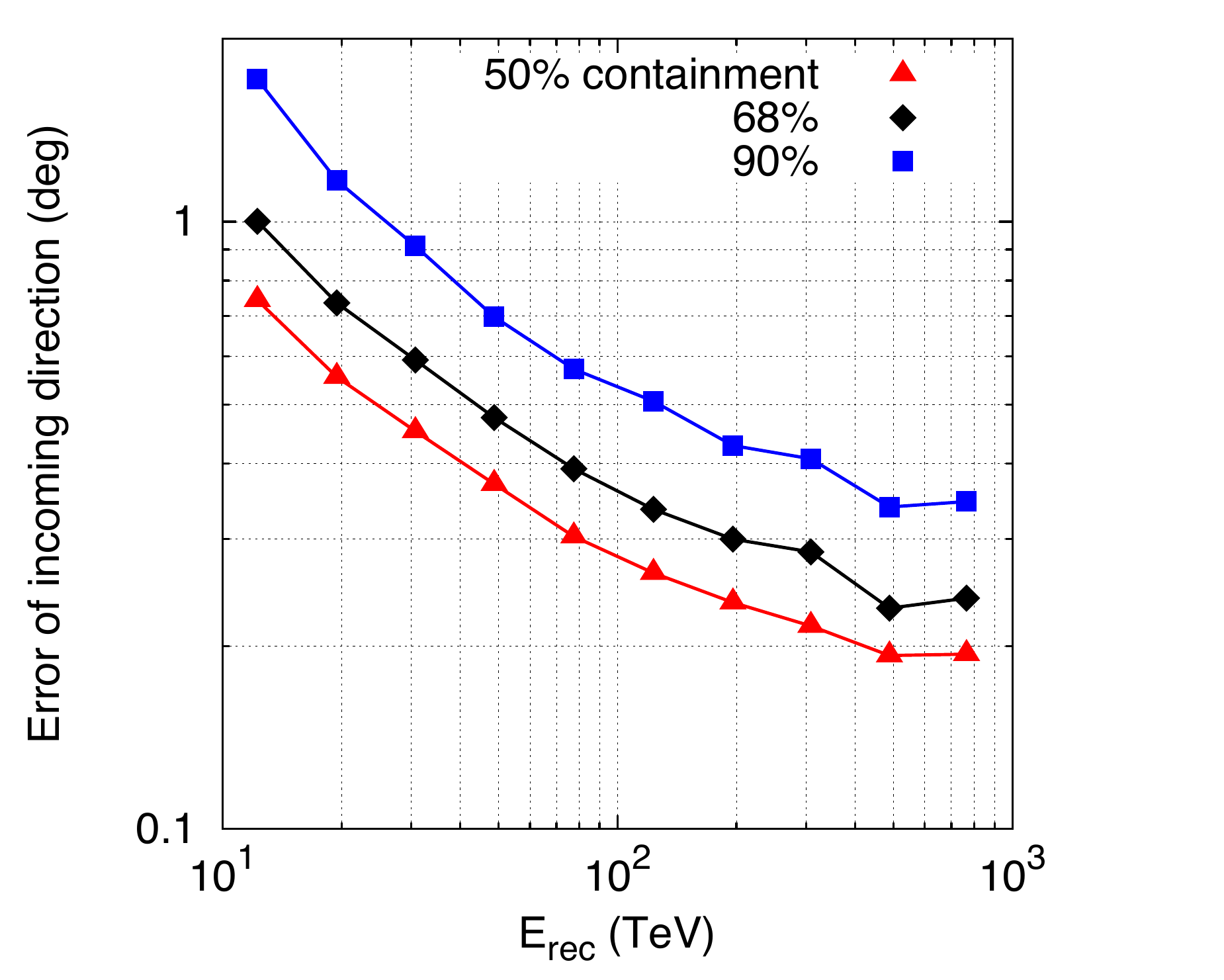}
  \end{center}
  \caption{Angular resolution (50\% containment, red triangle) of the ALPAQUITA AS array for primary gamma rays as a function of reconstructed energy. Also shown are the angular radii inside which 68\% (black diamond) and 90\% (blue square) of analyzed gamma-ray events are contained}
  \label{angular_resolution}
\end{figure}

Figure \ref{Core_res} shows the distance between the true shower core position and the reconstructed one with Equation \eqref{core_estimation}. The shower core position resolution is defined as the radius from the true core position inside which $50\%$ of reconstructed core positions are contained. The shower core position resolution improves monotonically until $E_{\rm rec} \simeq 100\, {\rm TeV}$ and then converges to $\simeq 2.5\, {\rm m}$. The angular resolution is defined in the same way as the shower core position resolution considering the accuracy of determination of direction shown in Fig. \ref{angular_resolution} (i.e., the angular radius inside which $50\%$ of analyzed events are contained). Table \ref{ALPAQUITA_AS_properties} summarizes the ALPAQUITA AS array performance in the reconstructed energy ranges of $10$, $50$, $100$, and $300\, {\rm TeV}$.
%As shown in Figure \ref{Ang_res}, the error of incoming direction has a different tendency from that of shower core position (Figure \ref{core_estimation}) in $E_{\rm rec} > 100\, {\rm TeV}$. This is because the shower events induced by gamma rays with larger energies have more scintillation directors used for the incoming direction reconstruction and improve the accuracy of detection timing of each detector, leading to a better angular resolution.

\begin{textblock*}{0.4\linewidth}(70pt, 520pt)
\begin{center}
  \begin{threeparttable}
    \caption{Performance of the ALPAQUITA AS array for primary gamma rays. $E_{\rm rec}$ means the reconstructed energy of gamma rays.}
    \begin{tabular}{cccc} \hline \hline
$E_{\rm rec}\, ({\rm TeV})$ & Energy resolution (\%) & Angular resolution ($^{\circ}$)\tnote{*} & Shower core position resolution (${\rm m}$)\tnote{*} \\ \hline
      10 & $+61 \, -38$ & 0.74 & 6.8\\ 
      50 & $+31 \, -27$ & 0.37 & 3.0\\
      100 & $+21 \, -23$ & 0.28 & 2.7\\
      300 & $+16 \, -18$ & 0.22 & 2.5\\ \hline
    \end{tabular}
    \label{ALPAQUITA_AS_properties}
    \begin{tablenotes}
    \item[*] 50\% containment.
    \end{tablenotes}
  \end{threeparttable}
\end{center}
\end{textblock*}
\clearpage
\section{Performance of the ALPAQUITA AS array + MD}\label{MDanalysis}
\subsection{Definition of the single muon and the number of muons per event in MD}
\begin{figure}[h]
  \begin{center}
    \includegraphics[scale=0.2]{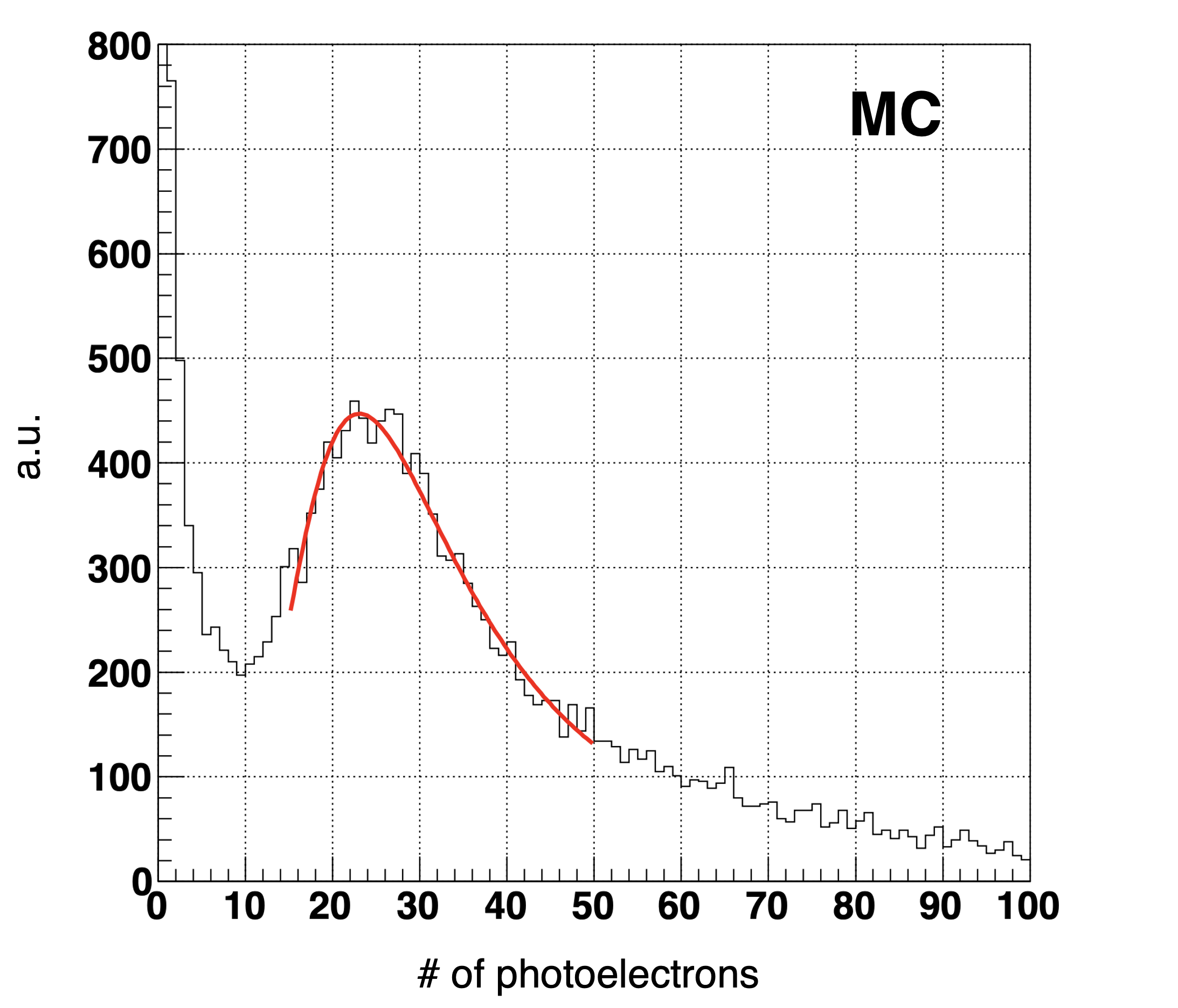}
  \end{center}
  \caption{Distribution of the number of photoelectrons recorded by an MD cell for CR events that trigger the AS array and have the reconstructed zenith angle within $30^{\circ}$. The red curve shows the best-fit result of the peak, assuming a Landau distribution}
  \label{1muon_def}
\end{figure}

For later analyses, it is convenient to define the single-particle signal in MD. Figure \ref{1muon_def} shows the distribution of the number of photoelectrons recorded by an MD cell, sampling CR showers. Assuming a Landau distribution, the distribution is fitted to find a peak at around 24 photoelectrons. The peak is regarded as being created by muons that go through MD and define the single-muon peak as 24 photoelectrons. This unit is employed in later analyses.

\subsection{Analysis conditions}\label{MD_Ana_conditions}
In the analysis using MD, the following two conditions are further imposed on reconstructed events in addition to conditions (1)-(3) described in Section \ref{Ana_conditions}: (4) the reconstructed zenith angle is within $40^{\circ}$ and (5) the position of a gamma-ray source is inside the analysis window opened for each event with an angular radius $r$ of
\begin{equation}\label{analysis_window}
  r = \begin{cases}
    1.5^{\circ} & (\Sigma \rho < 15\, {\rm m}^{-2}), \\
    \frac{5.8^{\circ}}{\sqrt{\Sigma \rho/{\rm m}^{-2}}} & (15\, {\rm m}^{-2} < \Sigma \rho < 135\, {\rm m}^{-2}), \, {\rm and} \\
    0.5^{\circ} & (135\, {\rm m}^{-2} < \Sigma \rho).
  \end{cases}
\end{equation}
The window radius is optimized so that it maximizes the quality factor (Q factor) of gamma rays after the selection.

\subsection{Event selection criterion using MD}\label{MD_ana}
\begin{figure}[h]
  \begin{center}
    \includegraphics[scale=0.22]{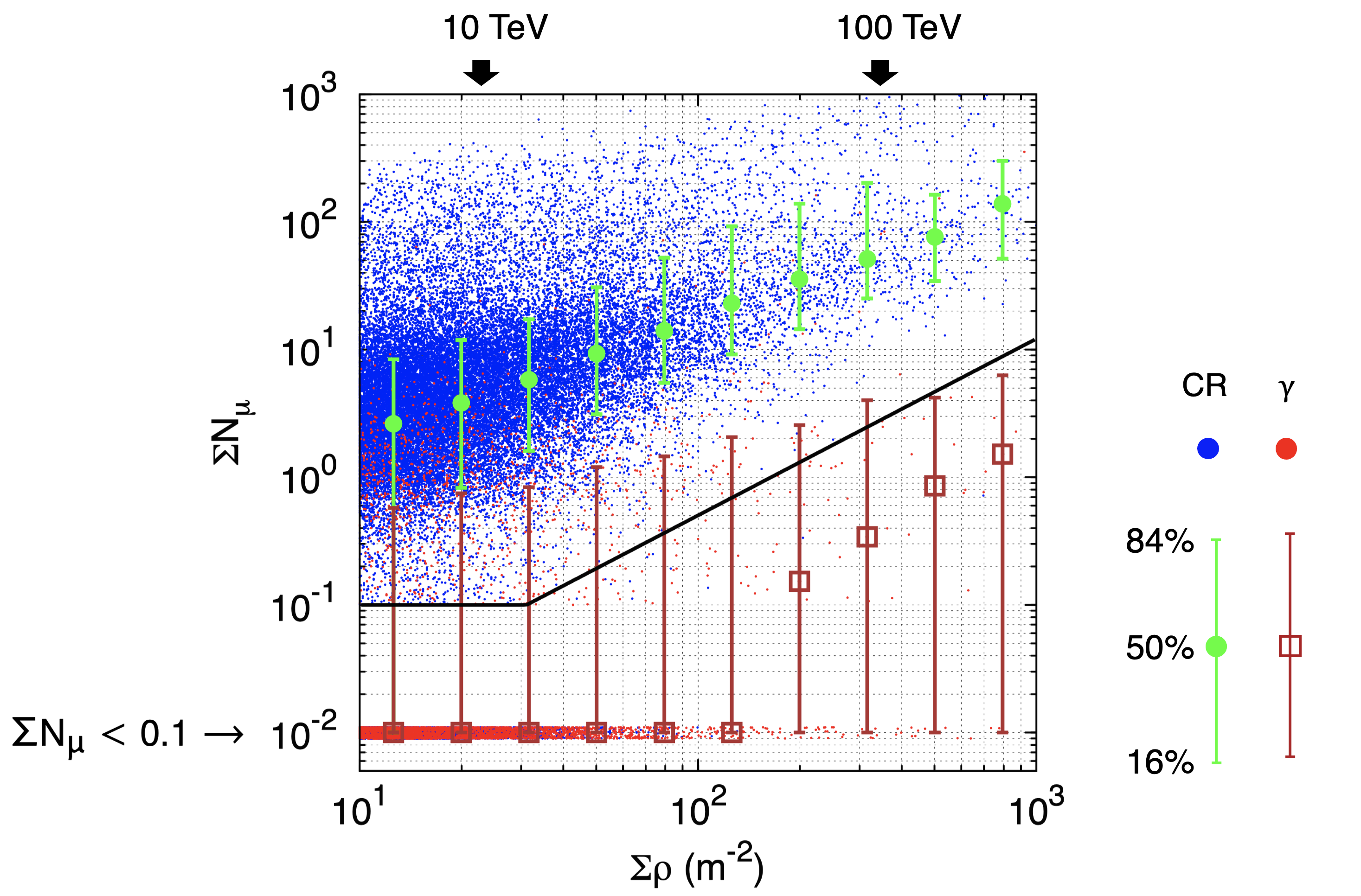}
  \end{center}
  \caption{Scatter plot of gamma-ray (red) and CR (blue) events in the ($\Sigma \rho$, $\Sigma N_{\mu}$) plane. Events with $\Sigma N_{\mu} < 0.1$ are piled up at around $\Sigma N_{\mu} = 0.01$ and the $\Sigma N_{\mu}$ values that contain 16\%, 50\%, and 84\% of analyzed events are shown by vertical bars both for gamma-ray (brown) and CR (light green) events. The thick black line shows the optimum $\Sigma N_{\mu}$ cut line used in the event selection criterion to discriminate between gamma-ray and CR events. Also shown are the corresponding gamma-ray equivalent energy ranges of $10\, {\rm TeV}$ and $100\, {\rm TeV}$ over the upper horizontal axis}
  \label{Scat_plot}
\end{figure}

Figure \ref{Scat_plot} shows a scatter plot of gamma-ray (red) and CR (blue) events in the ($\Sigma \rho$, $\Sigma N_{\mu}$) plane where $\Sigma N_{\mu}$ is defined as the total number of muons recorded by MD per shower event. In the analysis, the lower limit on the number of muons is defined as $0.1$ for all the MD cells, and events with $\Sigma N_{\mu} < 0.1$ are piled up at around $\Sigma \, N_{\mu} = 0.01$. Gamma-ray events are observed to be muon-poor, while CR events are muon-rich, and $\Sigma N_{\rm \mu}$ is useful to discriminate between them.

\begin{figure}[h]
  \begin{center}
    \includegraphics[scale=0.5]{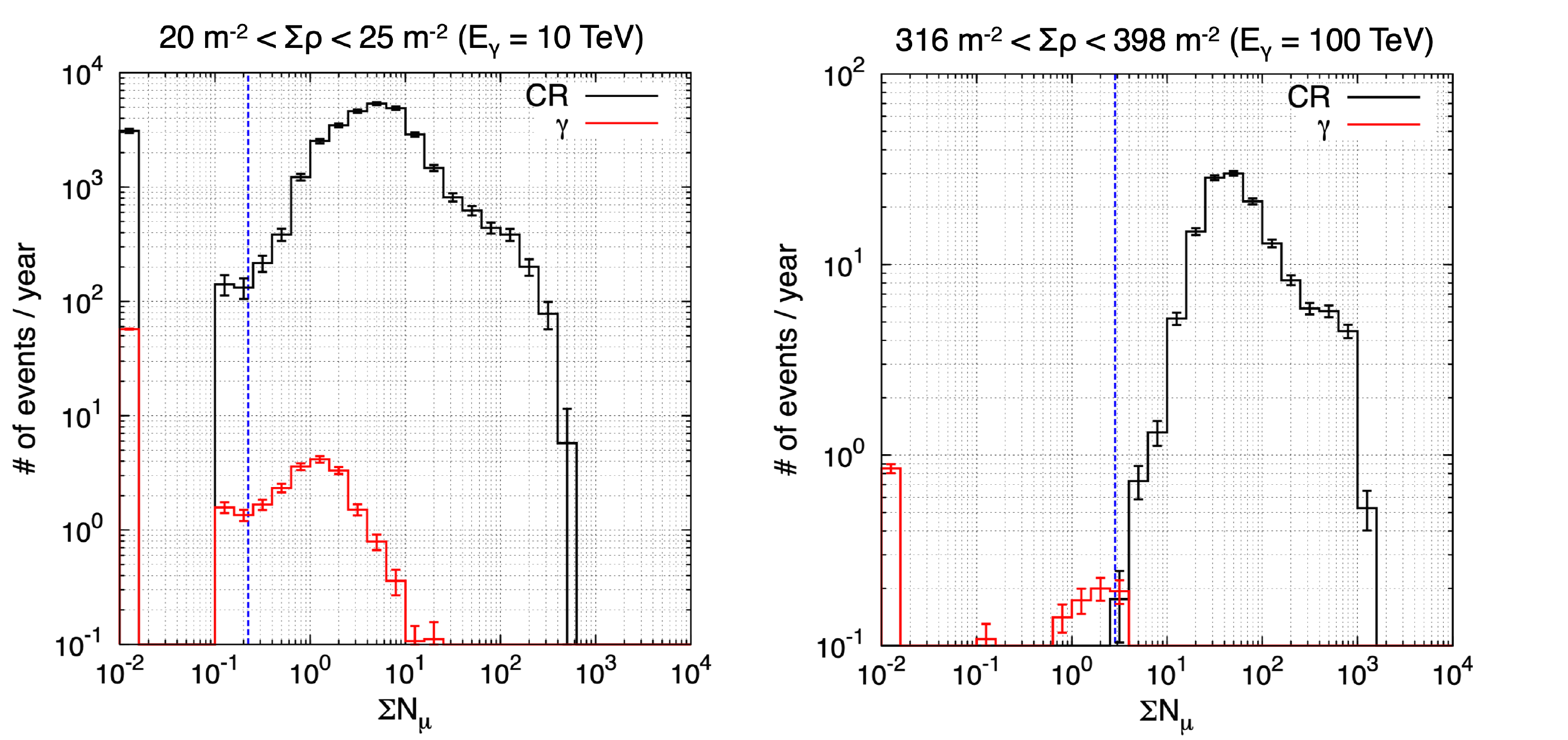}
  \end{center}
  \caption{Expected number of gamma-ray and CR events in one calendar year observation for $20\, {\rm m^{-2}} < \Sigma \rho < 25\, {\rm m^{-2}}$ corresponding to the gamma-ray equivalent energy $E_{\gamma}$ of $10\, {\rm TeV}$ (left) and $316\, {\rm m^{-2}} < \Sigma \rho < 398\, {\rm m^{-2}}$ ($E_{\gamma} = 100\, {\rm TeV}$, right). For the gamma rays, the energy spectrum of the Crab nebula modeled by F. Aharonian et al. (2004) \cite{HEGRA_Crab} is assumed. Events with $\Sigma N_{\mu} < 0.1$ are piled up at $\Sigma \, N_{\mu} = 0.01$. The dashed blue lines show the typical $\Sigma N_{\mu}$ cut value of each $\Sigma \rho$ bin (see the thick black line shown in Figure \ref{Scat_plot})}
  \label{Ev_distribution}
\end{figure}

The simulation data are normalized to the expected number of events corresponding to one calendar year observation. Figure \ref{Ev_distribution} shows the distribution of $\Sigma N_{\mu}$ both for gamma-ray and CR events. The distribution for gamma-ray events is calculated assuming the energy spectrum of the Crab Nebula modeled by F. Aharonian et al. (2004) \cite{HEGRA_Crab}:
\begin{equation}
  \frac{{\rm d}N}{{\rm d}E} = 2.83\times 10^{-11} \bigg(\frac{E}{\rm TeV}\bigg)^{-2.62} ({\rm cm}^{-2}\, {\rm s}^{-1}\, {\rm TeV}^{-1}).\nonumber
\end{equation}
For the background CR events, taking into account its isotropic characteristics, the number of events in the simulation is converted as
\begin{equation}
  F_{\rm CR}T_{1{\rm yr}, {\theta}<60^{\circ}} S_{\rm sim} \frac{\Omega\sum_{i}^{N}{r_i}^2}{N_{\rm CR, all}},\nonumber
\end{equation}
where $F_{\rm CR} = 1.03\times 10^{-4}\, ({\rm cm}^{-2}\, {\rm s}^{-1}\, {\rm sr}^{-1})$ is the integral flux of CRs including all species above $300\, {\rm GeV}$, $T_{1{\rm yr}, {\theta}<60^{\circ}}$ is the duration of the hypothetical source being within the zenith angle range smaller than $60^{\circ}$ at the ALPAQUITA site, $S_{\rm sim} = 2.83\times 10^{9}\, ({\rm cm}^{2})$ is the area within which shower cores are distributed in the simulation, $\Omega = 9.57\times 10^{-4}\, ({\rm sr})$ is the solid angle of a circular region with an apparent radius of $1^{\circ}$, $r_i\, (^{\circ})$ is angular radius of the analysis window defined by Equation \eqref{analysis_window} opened for each reconstructed event, and $N_{\rm CR, all}$ is the total number of cosmic-ray events generated with CORSIKA. The summation is taken over all the reconstructed events that pass through the analysis conditions (1) to (4) described in Section \ref{Ana_conditions} and \ref{MD_Ana_conditions}.
%for $25.1\, {\rm m}^{-2}\le \Sigma \rho < 39.8\, {\rm m}^{-2}$ ($E_{\rm rep} \simeq 10\, {\rm TeV},\,  {\rm left}$) and $251\, {\rm m}^{-2}\le \Sigma \rho < 398\, {\rm m}^{-2}$ ($E_{\rm rep} \simeq 100\, {\rm TeV},\, {\rm right}$)

\begin{figure}[h]
  \begin{center}
    \includegraphics[scale=0.5]{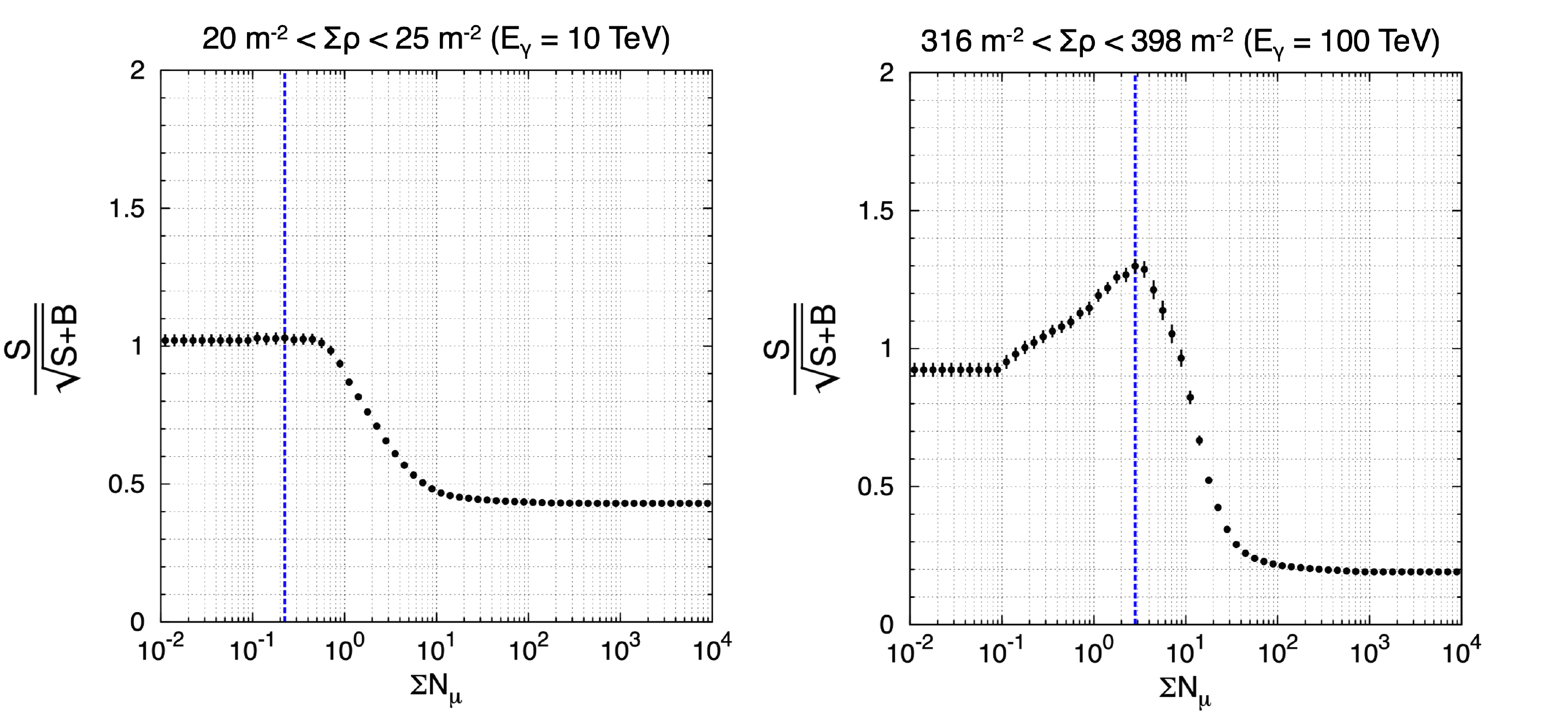}
  \end{center}
  \caption{Q-factor distributions for two $\Sigma \rho$ ranges; $20\, {\rm m^{-2}} < \Sigma \rho < 25\, {\rm m^{-2}}$ ($E_{\gamma} = 10\, {\rm TeV}$, left) and $316\, {\rm m^{-2}} < \Sigma \rho < 398\, {\rm m^{-2}}$ ($E_{\gamma} = 100\, {\rm TeV}$, right). The vertical axis indicates the Q factor $S/\sqrt{S+B}$ ($S$ the number of gamma-ray events, and $B$ that of CR events) calculated by taking events with $\Sigma N_{\mu}$ smaller than a fixed cut value shown in the horizontal axis. The dashed blue lines show the typical $\Sigma N_{\mu}$ cut value of each $\Sigma \rho$ bin as the same as in Figure \ref{Ev_distribution}}
  \label{FOM_figure}
\end{figure}

\begin{figure}[h]
  \begin{center}
    \includegraphics[scale=0.3]{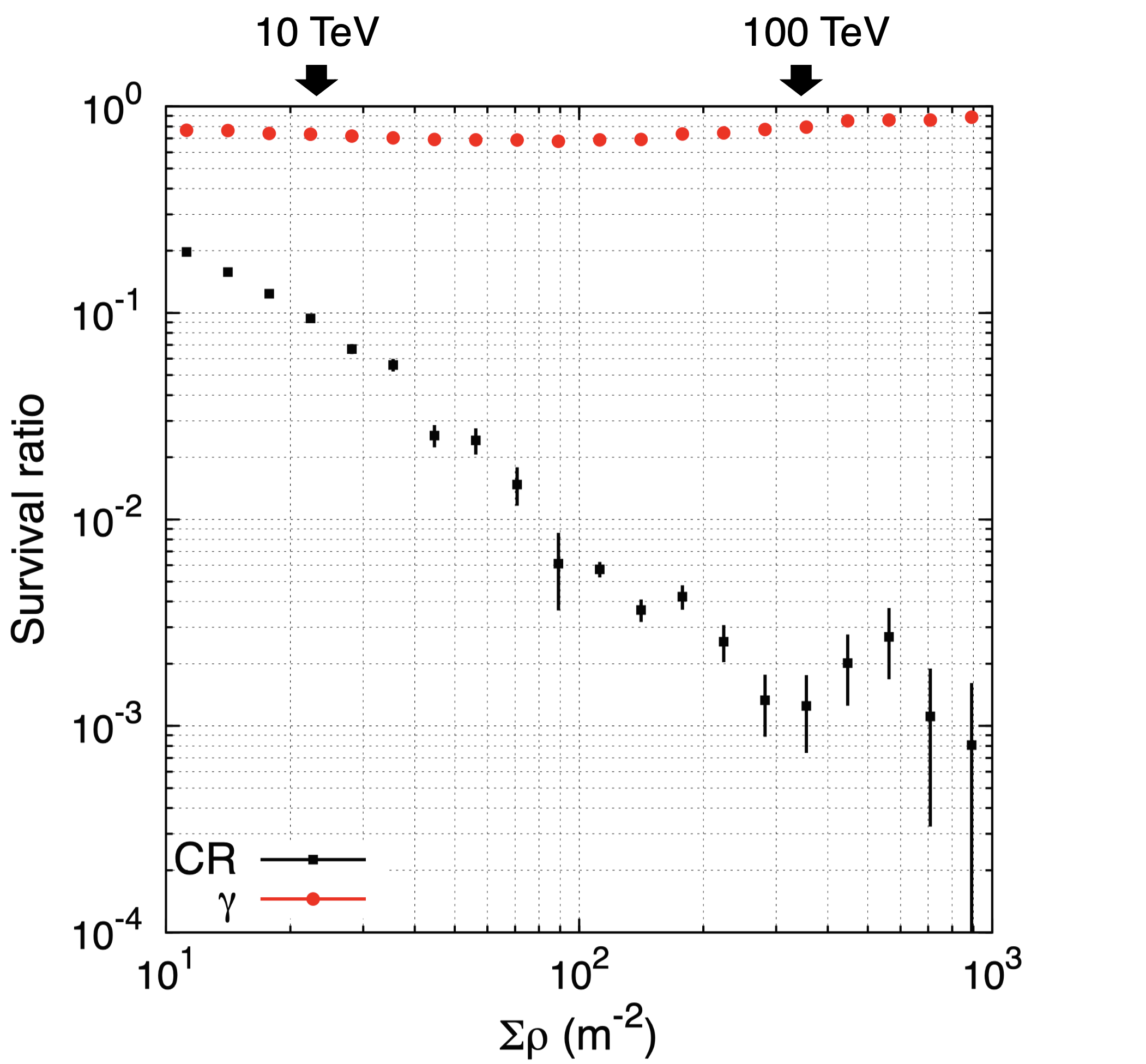}
  \end{center}
  \caption{Survival ratio of gamma-ray (red) and CR (black) events after applying the $\Sigma N_{\mu}$ selection criterion (see the thick black line shown in Figure \ref{Scat_plot}). Also shown are the corresponding gamma-ray equivalent energy ranges of $10\, {\rm TeV}$ and $100\, {\rm TeV}$ over the upper horizontal axis}
  \label{Survival_effi}
\end{figure}

Figure \ref{FOM_figure} shows the Q-factor distributions in two $\Sigma \rho$ ranges corresponding to the gamma-ray equivalent energy ranges of $10\, {\rm TeV}$ (left) and $100\, {\rm TeV}$ (right). Taking the Q factors of all the analyzed $\Sigma \rho$ bins into account, the optimum cut line is determined as a function of $\Sigma \rho$, shown by the thick black line in Figure \ref{Scat_plot}. Figure \ref{Survival_effi} shows the survival ratio of gamma-ray (red) and CR (black) events after applying the $\Sigma N_{\mu}$ selection criterion. The error bar of the point in $100\le \Sigma \rho \le 126$ is smaller than that of the point in $79\le \Sigma \rho \le 100$, because both the CR simulation data generated in $300\, {\rm GeV} \le E \le 10\, {\rm PeV}$ and in $10\, {\rm TeV} \le E \le 10\, {\rm PeV}$ are used above $\Sigma \rho \ge 100$. Beyond $100\, {\rm TeV}$, high rejection power of CRs ($\simeq 99.9\%$) is achieved. The expected number of CR events that contaminate gamma-ray events is $0.70 \pm 0.14$ per calendar year for a point source above the gamma-ray equivalent energy range of $100\, {\rm TeV}$. For gamma-ray events, high survival ratios ($\gtrsim 80\%$) are achieved beyond $100\, {\rm TeV}$ after applying the $\Sigma N_{\mu}$ selection criterion.

\subsection{Sensitivity to southern gamma-ray sources}\label{sensitivity}
\begin{figure}[h]
  \begin{center}
    \includegraphics[scale=0.5]{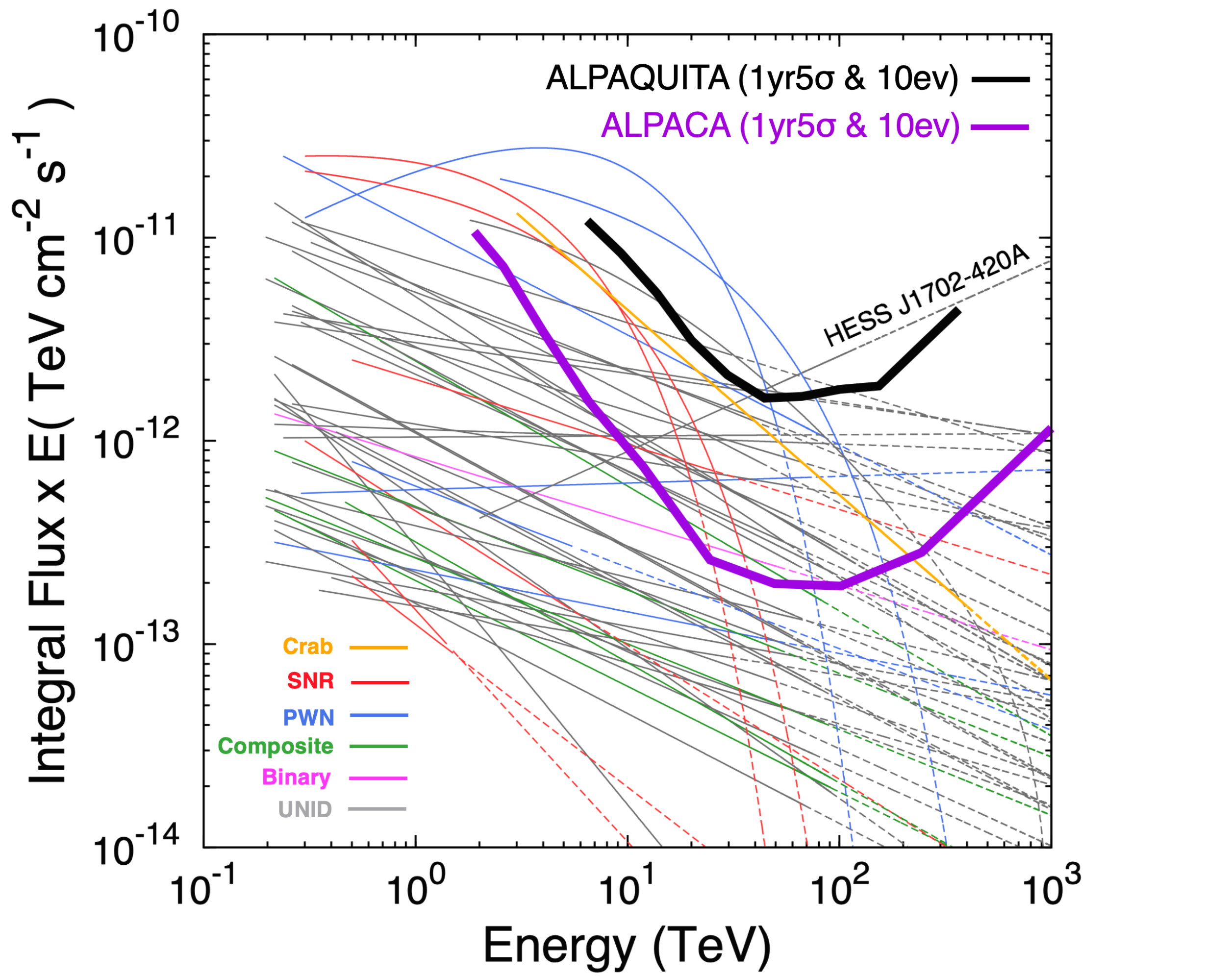}
  \end{center}
  \caption{Sensitivity curve of ALPAQUITA (the thick black curve) for a gamma-ray point source together with the energy spectra of the H.E.S.S. \cite{HGPS, 1702AandB} and HAWC \cite{HAWC_56TeV} gamma-ray sources that are in the ALPAQUITA field of view. The thick purple curve shows the estimated sensitivity of ALPACA. The ALPACA curve is derived by scaling the sensitivity curve of Tibet AS$\gamma$ \cite{TibetMD} considering the ratio of the areas of these two experiments. Regarding the energy spectra, different colors indicate different source species: supernova remnants (SNR) in red, pulsar wind nebulae (PWN) in blue, composite SNRs (Composite) in green, compact binary systems (Binary) in magenta and unidentified sources (UNID) in gray, respectively. The Crab Nebula spectrum modeled by M. Amenomori et al. (2019) \cite{tibet_100TeVCrab} is shown in orange. Solid and dashed lines show observed and extrapolated regions, respectively. In extrapolating the spectra, the attenuation of gamma rays due to the $e^{+}e^{-}$ pair production with the interstellar radiation field is not taken into consideration}
  \label{Sensitivity}
\end{figure}

Figure \ref{Sensitivity} shows the sensitivity curve of ALPAQUITA (the thick black curve) together with the energy spectra of the H.E.S.S. and eHWC gamma-ray sources \cite{HAWC_56TeV, HGPS, RXJ1713, HESSJ1911+090, HESSJ1718-374, HESSJ1731-347, HESSJ1747-281, 1702AandB} that are in the ALPAQUITA field of view. The sensitivity is for a point source with $5\, \sigma$ significance detection in one calendar year observation. In the energy region where the expected number of background CR events per calendar year is smaller than 1, the number of gamma rays is required to exceed $10$. According to Figure \ref{Sensitivity}, five sources $-$ HESS J0835-455, HESS J1825-137, HESS J1908+063, HESS J1616-508, and HESS J1702-420A $-$ will be detected beyond $10\, {\rm TeV}$, and the detection of gamma-rays beyond $100\, {\rm TeV}$ is also possible for HESS J1702-420A if the spectrum extends without cutoff.

For the extrapolated parts of the gamma-ray spectra, the attenuation caused by the interstellar background photons is not considered. According to Vernetto and Lipari (2016) \cite{Gamma_attenuation}, gamma rays of $\simeq 150\, {\rm TeV}$ coming from the Galactic Center can be attenuated by $\simeq 20\%$ until reaching our solar system due to the $e^{+}e^{-}$ pair production with infrared photons thermally emitted from the interstellar dust. Gamma rays of $\simeq 1\, {\rm PeV}$ can also be attenuated by $\simeq 70\%$ due to the interaction with the CMB photons.

\begin{figure}[h]
  \begin{center}
    \includegraphics[scale=0.5]{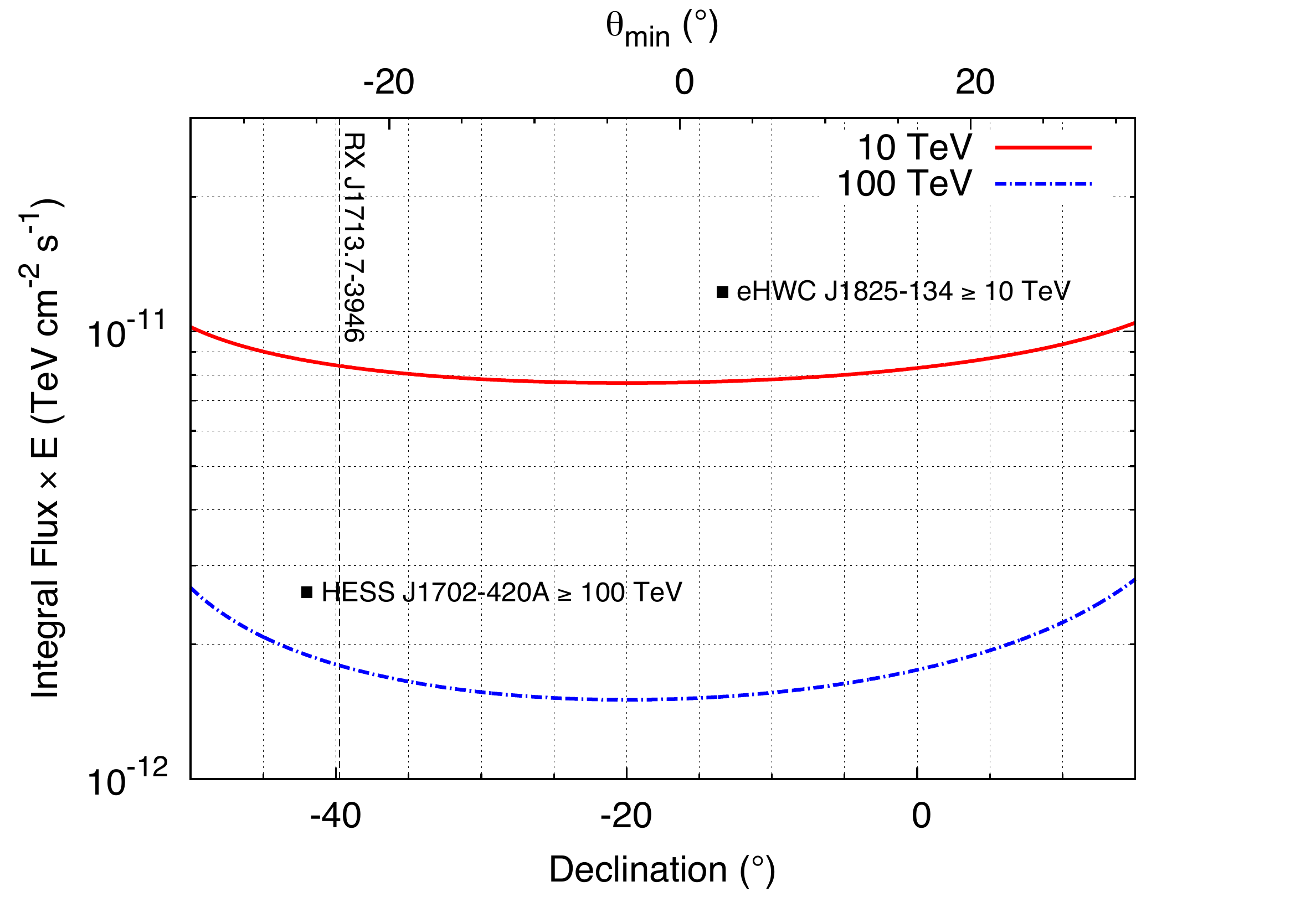}
  \end{center}
  \caption{Effect of the declination dependence of exposure on the ALPAQUITA sensitivity to gamma-ray sources above $10\, {\rm TeV}$ (solid red line) and $100\, {\rm TeV}$ (dotted-dashed blue line). The lower horizontal axis shows the declination of a gamma-ray source, while the upper horizontal axis the minimum zenith angle which a source in a fixed declination reaches at the ALPAQUITA site. The vertical dashed line shows the declination of RX J1713.7-3946, whose path in the sky is assumed in the air shower simulation (see Section \ref{CORSIKA}). Also shown by black squares are the integral fluxes of eHWC J1825-134 \cite{HAWC_56TeV} and HESS J1702-420A \cite{1702AandB} above $10\, {\rm TeV}$ and $100\, {\rm TeV}$, respectively. For both two fluxes, the attenuation of gamma rays due to the interaction with the interstellar radiation field is not considered. The horizontal axis of each square indicates the declination of the corresponding source}
  \label{exposure}
\end{figure}

Furthermore, since the ALPAQUITA sensitivity curve is derived on the basis of detector response to a hypothetical Crab-like source on the path in the sky of RX J1713.7-3946 (see Section \ref{CORSIKA} and \ref{MD_ana}), the declination dependence of exposure should be taken into account to properly evaluate the detectability of each gamma-ray source. Figure \ref{exposure} shows how much the declination dependence of exposure affects ALPAQUITA sensitivity above $10\, {\rm TeV}$ and $100\, {\rm TeV}$. The path in the sky assumed in this study does not lead to the best sensitivity of ALPAQUITA. However, taking into account factors such as gamma-ray absorption by the interstellar background photons and the declination dependence of exposure does not affect the conclusion about the source detection.

\paragraph{HESS J1702-420A} HESS J1702-420A is a gamma-ray point source discovered by H.E.S.S. along with the surrounding extended source HESS J1702-420B \cite{1702AandB}. The relation between these two sources is not clear. The energy spectra of both sources extend up to $\simeq 100\, {\rm TeV}$ without showing cutoff, and HESS J1702-420A dominates the total gamma-ray flux beyond $50\, {\rm TeV}$ with its extremely hard spectral index ($\simeq 1.5$). Although SNR G344.7-0.1 and PSR J1702-4128 are in the vicinity of the gamma-ray emission region, it is not easy to consider these objects as the origin of the emission \cite{MNRAS_1702, HESS_1702_again}. The absence of X-ray flux \cite{Xray_1702, Xray_1702_Suzaku} and the observation of gamma rays in $10\, {\rm GeV}$ to $30\, {\rm TeV}$ \cite{Eagle} do not favor the leptonic origin scenario of the VHE gamma-ray emission, but the hadronic scenario is not conclusive because of the lack of clear correlation between the VHE gamma-ray emission region and the ISM distribution \cite{1702AandB, Giacani}. According to Figures \ref{Sensitivity} and \ref{exposure}, ALPAQUITA will detect HESS J1702-420A above $\simeq 300\, {\rm TeV}$ with its one calendar year observation if the spectrum extends without cutoff and to provide data to discuss the mechanism of the particle acceleration taking place in this peculiar object.

\section{Conclusion}\label{Conclusion}
Gamma-ray observation beyond $100\, {\rm TeV}$ is crucial to specify the origin of CRs around the knee energy region. ALPACA is an international experiment for cosmic-ray physics in the southern hemisphere that aims to elucidate this long-standing mystery. The ALPAQUITA experiment has been designed as the prototype experiment of ALPACA. Data acquisition is expected to start in late 2021. ALPAQUITA consists of a surface AS array composed of plastic scintillation detectors of $1\, {\rm m}^2$ area and an underground MD. The AS array and MD have total areas of $18,450\, {\rm m}^2$ and $900\, {\rm m}^2$, respectively, and will be expanded to the ALPACA arrays shortly. A Monte Carlo simulation evaluates the ALPAQUITA performance, including sensitivity to gamma-ray sources. ALPAQUITA achieves 100\% trigger efficiency above $20{\rm TeV}$. Using the reconstructed energy $E_{\rm rec}$ of primary gamma rays, the upper and lower energy resolutions are empirically formulated as $\simeq 21\, (0.5\, {\rm log}_{10}(E_{\rm rec}/100\, {\rm TeV}) + 1)^{-1.5}\, \%$ and $\simeq 23\, (0.5\, {\rm log}_{10}(E_{\rm rec}/100\, {\rm TeV}) + 1)^{-0.7}\, \%$ in $10\, {\rm TeV} < E_{\rm rec} < 1000\, {\rm TeV}$, respectively. The angular resolution also reaches $\simeq 0.3^{\circ}$ at $E_{\rm rec} = 100\, {\rm TeV}$ thanks to the accurate estimation of shower core position with the resolution of $2.7\, {\rm m}$.

Note that several optimization procedures have yet to be carried out. For example, the power of the weights used in the core position estimation (currently set at 2 in Equation \eqref{core_estimation}) and the slope of the conical fit ($b$ in Equation \eqref{cone_fit}) are not yet optimized for ALPAQUITA. Fitting of the NKG function to an observed lateral shower distribution also leads to a better energy resolution \cite{S50} compared to the method employed in this paper. These optimizations will be performed shortly.

After applying the $\Sigma N_{\mu}$ selection criterion using MD, $\simeq 80\%$ of gamma-ray events survive beyond $100 \, {\rm TeV}$ while $\simeq 99.9\%$ of CR events are rejected. The expected number of CR events that contaminate gamma-ray events from a point source is less than one above the gamma-ray equivalent energy range of $100\, {\rm TeV}$ in one calendar year observation. The sensitivity of ALPAQUITA to gamma-ray point sources is calulated to find that five sources will be detected above $5\, \sigma$ significance in one calendar year observation beyond $10\, {\rm TeV}$, with one out of the five $-$ HESS J1702-420A $-$ above $\simeq 300\, {\rm TeV}$. HESS J1702-420A does not have any clear counterparts in other wavelength ranges and will be the case in point to discuss the emission mechanism of gamma rays beyond $100\, {\rm TeV}$ with ALPAQUITA in its brief operation as the prototype of ALPACA.

\section*{Acknowledgements}
The ALPACA project is supported by the Japan Society for the Promotion of Science (JSPS) through Grants-in-Aid for Scientific Research (A) 19H00678, Scientific Research (B) 19H01922, and Scientific Research (S) 20H05640, the LeoAtrox supercomputer located at the facilities of the \textit{Centro de Análisis de Datos (CADS)}, CGSAIT, Universidad de Guadalajara, México, and by the joint research program of the Institute for Cosmic Ray Research (ICRR), The University of Tokyo. K. Kawata is supported by the Toray Science Foundation. E. de la Fuente thanks Coordinación General Académica y de Innovación (CGAI-UDG), cuerpo académico PRODEP-UDG-CA-499, Carlos Iván Moreno, Cynthia Ruano, Rosario Cedano, and Diana Naylleli, for financial and administrative support during sabbatical year stay at the ICRR on 2021. I. Toledano-Juarez acknowledges support from CONACyT, México; grant 754851. F. Orozco-Luna thanks CONACyT Ph. D. Grant 2021-000001-01NACF-02328.

\clearpage
%\begin{comment}

%\end{comment}
\clearpage
\bibliographystyle{spphys}
\bibliography{mybibfile}

\end{document}